 \documentclass[final,5p,times,twocolumn,authoryear]{elsarticle}

\usepackage{amssymb}
\usepackage{lipsum}
\usepackage{amsmath,amssymb,amsfonts,dcolumn,color,graphicx,graphics,latexsym,placeins,epsfig}
\usepackage{epsfig}
\usepackage{bm}
\usepackage{slashed}
\usepackage{latexsym}
\usepackage{natbib}
\usepackage{url}
\usepackage{dcolumn}
\usepackage{color}
\usepackage{amsfonts,amssymb,amsmath}
\usepackage{graphicx,epsfig}
\usepackage{psfrag}
\usepackage{subfigure}
\usepackage{tabularx}
\usepackage{comment}

\journal{Physics Letters B}

\begin{document}

\begin{frontmatter}



\title{Primordial gravitational waves from spontaneous Lorentz symmetry breaking}


\author[first]{Mohsen Khodadi}
\ead[first]{khodadi@kntu.ac.ir}
\affiliation[first]{organization={School of Physics, Institute for Research in Fundamental Sciences (IPM)},
            city={Tehran},
            postcode={P.O. Box 19395-5531}, 
            country={Iran}}

\author[second]{Gaetano Lambiase}
\ead[second]{lambiase@sa.infn.it}
\affiliation[second]{organization={ Dipartimento di Fisica ``E.R. Caianiello", Università di Salerno},
       addressline={Via Giovanni Paolo II},
            city={Fisciano},
            postcode={I-84084}, 
            country={Salerno} }

\author[third]{Leonardo Mastrototaro}
\ead[third]{lmastrototaro@unisa.it}
\affiliation[third]{organization={ Dipartimento di Fisica ``E.R. Caianiello", Università di Salerno},
       addressline={Via Giovanni Paolo II},
            city={Fisciano},
            postcode={I-84084}, 
            country={Salerno}}       

\author[fourth]{Tanmay Kumar Poddar}
\ead[fourth]{poddar@sa.infn.it}
\affiliation[fourth]{organization={INFN, Gruppo collegato di Salerno},
            addressline={Via Giovanni Paolo II}, 
            city={Fisciano},
            postcode={I-84084}, 
            state={Salerno},
            country={Italy}}

\begin{abstract}
We study the effect of Spontaneous Lorentz Symmetry Breaking (SLSB) on Primordial Gravitational Waves (PGWs) generated during inflation. The SLSB is induced by a time-like Bumblebee vector field which is non-minimally coupled to the Ricci tensor in the Friedmann-Lema\^itre-Robertson-Walker background.
		The power spectrum and GW amplitude are computed to investigate how Lorentz violation leaves observable imprints. We calculate the GW strain amplitude over frequencies \((10^{-10}~\mathrm{Hz}, 10^4~\mathrm{Hz})\), for a range of the dimensionless Lorentz-violating parameter, \( -10^{-3} \leq l \leq 10^{-4} \), which essentially comes from a slight sensitivity to the equation of state for dark energy. For positive \( l \) values, the amplitude of GW shows a mild suppression compared to the standard cosmological scenario (\( l = 0 \)). This effect could be observable with detectors like SKA, \(\mu\)-Ares, and BBO. Conversely, negative \( l \) values amplify the GW amplitude, enhancing detectability by both SKA, $\mu$-Ares, and BBO, as well as by THEIA and DECIGO. Notably, the GW strain amplitude increases by an order of magnitude as \( l \) moves from 0 to \( -10^{-3} \), improving prospects for detection in high-sensitivity detectors like THEIA and DECIGO.
\end{abstract}




\end{frontmatter}




\section{Introduction}
\label{sec1}

Primordial Gravitational Waves (PGWs) are spacetime ripples originating from the early universe, shortly after the Big Bang \citep{Maggiore:1999vm,LIGOScientific:2019vic,Wang:2024gko}. A key source of PGWs is cosmic inflation—a brief period of rapid exponential expansion that addresses major cosmological puzzles such as the horizon, flatness, and monopole problems, along with structure formation and observed Cosmic Microwave Background (CMB) anisotropies \citep{Guth:1980zm,Achucarro:2022qrl,Kamionkowski:1996ks,Turner:1996ck,Zaldarriaga:1996xe}.
	
	Inflation occurred over a tiny timescale, roughly \( 10^{-36}~\mathrm{s} \) to \( 10^{-32}~\mathrm{s} \), during which the universe expanded by at least a factor of \( e^{60} \approx 10^{26} \) \citep{Achucarro:2022qrl}. Quantum vacuum fluctuations in the gravitational field during this era induced spacetime perturbations. These fluctuations were stretched beyond the Hubble horizon, freezing into classical perturbations as their wavelengths exceeded the observable universe.
	
	These tensor perturbations persisted as PGWs and re-entered the horizon during later cosmic evolution. Inflationary models predict a nearly scale-invariant power spectrum for PGWs, with a slight red tilt characterized by the spectral index (\( n_T \)), linked to the slow-roll parameter of the inflationary potential \citep{Guzzetti:2016mkm,Kamionkowski:2015yta,Vagnozzi:2023lwo,LISACosmologyWorkingGroup:2024hsc}.
	
	The amplitude (\( h \)) of PGWs depends on the inflationary energy scale, with higher scales producing stronger signals. Key characteristics of PGWs include the tensor-to-scalar ratio (\( r \)), which measures the relative strength of tensor perturbations (PGWs) to scalar perturbations (density fluctuations), and the spectral index. These parameters offer valuable insights into the dynamics of inflation and the physics of the early universe \citep{Achucarro:2022qrl}. The combined analysis of data from Planck, BICEP/Keck 2018, and Baryon Acoustic Oscillation (BAO) measurements places a stringent upper limit on the tensor-to-scalar ratio, with \( r < 0.032 \) at a $95\%$ confidence level \citep{Tristram:2021tvh}.

PGWs can also originate from non-inflationary processes, such as first-order phase transitions in the early universe, including the electroweak \citep{Apreda:2001us,Leitao:2015fmj,Shajiee:2018jdq,Goncalves:2021egx,Mohamadnejad:2021tke,Weir:2017wfa} and Quantum Chromodynamics (QCD) \citep{Caprini:2010xv,Aoki:2017aws,Anand:2017kar,Brandenburg:2021tmp,Ahmadvand:2017tue,Ahmadvand:2017xrw,Chen:2017cyc,Cutting:2018tjt,Capozziello:2018qjs,Khodadi:2018scn,Li:2018oqf,Shakeri:2018qal,Hajkarim:2019csy,Davoudiasl:2019ugw,Dai:2019ksi,Rezapour:2020mvi,Khodadi:2021ees,Feng:2022fwf} epochs, from cosmic strings \footnote{While cosmic strings have remained undetected in conventional observations, the advent of GW astronomy could soon change that. Their signature may finally be revealed through the PGW background. Intriguingly, recent pulsar timing array data has uncovered a nanohertz-frequency stochastic GW signal, with cosmic strings emerging as a possible source \citep{EPTA:2023xxk}. A cosmic string network’s PGW offers a probe of both the phase transition energy scale and the high-energy particle physics responsible for their existence (for a review see Ref. \citep{Sousa:2024ytl}).} \citep{Battye1998,Chang:2021afa}, Primordial Black Holes (PBHs) \citep{Sasaki:2018dmp} and in different modified gravity theories such as \citep{Mavromatos:2022yql,Oikonomou:2022ijs,Jizba:2024klq,Odintsov:2024sbo,Odintsov:2022cbm}.

	PGWs leave measurable imprints on the CMB, including anisotropies and the characteristic B-mode polarisation pattern. They also affect the universe's energy density during Big Bang Nucleosynthesis (BBN), influencing the expansion rate and altering the abundances of light elements. Comparing predicted and observed abundances provide constraints on the energy density of PGWs \citep{Clarke:2020bil}. The CMB+BBN observations put limit on the GW energy density as $\Omega_{\mathrm{GW}}< 1.5\times 10^{-5}$ for frequency $1.8\times 10^{-11}~\mathrm{Hz}<f< 4.5\times 10^{8}r^{1/4}~\mathrm{Hz}$ \citep{Boyle:2007zx}, if it is sourced by inflation.
	
	The sources and power spectrum of PGW discussed above are based on Einstein's General Relativity (GR) and the cosmological principle, which assumes spacetime homogeneity and isotropy. However, both GR and the SM of particle physics are low-energy effective theories, potentially arising from a more fundamental high-energy framework. Any significant deviations observed in precision measurements could signal violations of the fundamental symmetries of these theories. It is well established that Lorentz symmetry ensures the speed of light in a vacuum is independent of its source. However, investigations into gravitational radiation indicate a potential violation of Lorentz symmetry, which could result in a source-dependent velocity of GWs \citep{Ghosh:2023xes}. This would significantly influence their propagation over cosmic distances.
	
	This paper explores the effects of Spontaneous Lorentz Symmetry Breaking (SLSB), a promising modification in early-universe cosmology, on the standard PGW spectrum. 
	
	Early discussions on Lorentz Invariance Violation (LIV) trace back to foundational works \citep{Dirac1951,Bjorken:1963vg,Pavlopoulos:1967dm}. Experimentally, LIV's significance lies in its potential to produce minute but measurable effects well below the Planck scale, enabled by advances in precision techniques. This underscores the need for a robust theoretical and phenomenological framework.
	
	The Standard Model Extension (SME) is a widely established, model-independent effective field theory that incorporates LIV into high-energy particle phenomenology \citep{Colladay:1996iz,Colladay:1998fq}, parametrising it with 19 independent coefficients. The SME has also been extended to gravity, where Lorentz Symmetry Breaking (LSB) is formulated to preserve geometric constraints and conservation laws \citep{Kostelecky:2003fs} (see also \citep{Bluhm:2004ep,Kostelecky:2010ze}).
	
	Bumblebee Gravity (BG) provides a simpler alternative to the SME for incorporating LIV. In BG, the gravity couples to a bumblebee vector field with a nonzero Vacuum Expectation Value (VEV) \citep{Kostelecky:2003fs,Kostelecky:2010ze,Maluf:2021lwh} \footnote{Other LIV-based theories include string theory \citep{Kostelecky:1988zi,Kostelecky:1989jw}, noncommutative spacetime \citep{Carroll:2001ws}, Aether gravity \citep{Eling:2004dk}, Horava-Lifshitz gravity \citep{Horava:2009uw}, and Mimetic gravity \citep{Chamseddine:2013kea}. For a review of Lorentz symmetry breaking in curved spacetime, see Ref. \citep{Mariz:2022oib}}. BG extends GR by incorporating SLSB, offering a promising framework for addressing various aspects of cosmology and gravity which sheds light on our understanding of fundamental physics \citep{Jesus:2019nwi,Jesus:2020lsv,Santos:2014nxm,Capelo:2015ipa,Neves:2022qyb,Valtancoli:2023kdy,Capozziello:2023rfv,Capozziello:2023tbo,Ovgun:2018xys,Maluf:2020kgf,Gullu:2020qzu,Delhom:2019wcm,Ding:2019mal,Jha:2020pvk,Filho:2022yrk,Xu:2022frb,Ding:2023niy,Mai:2024lgk,AraujoFilho:2024ykw,Liu:2024axg,Ding:2024qrf,Neves:2024ggn,Ji:2024aeg,AraujoFilho:2025hkm,Vagnozzi:2022moj}.
	
	A no-go theorem \citep{Kostelecky:2003fs} demonstrates that explicit LIV with fixed, non-dynamical background fields violates the geometric constraints and conservation laws of GR. This issue is resolved by SLSB, analogous to the Higgs mechanism. SLSB, first proposed in bosonic string field theory, arises in high-energy regimes through a potential term in the Lagrangian that induces a vacuum state with nonzero VEVs.
	
	The BG model incorporates a nonzero VEV, indicating SLSB through a dimensionless parameter \( l \), typically positive and constrained by small observational limits \citep{Bertolami:2005bh,Paramos:2014mda,Casana:2017jkc}. However, studies of Black Holes (BHs) and Neutron Stars (NSs) suggest \( l \) could be negative, with weaker constraints compared to positive values \citep{Khodadi:2021owg,Khodadi:2022dff,Khodadi:2023yiw, Wang:2021gtd,Jha:2021eww}. While most constraints are astrophysical, limited cosmological studies \citep{Maluf:2021lwh,Khodadi:2022mzt} emphasize the need for further exploration.

	This manuscript models Friedmann-Lema\^itre-Robertson-Walker (FLRW) cosmology within BG, exploring its impact on the early universe's dynamics. We examine how SLSB from a Bumblebee vector field in an FLRW background affects the strain amplitude of PGWs generated during the inflationary phase. In more complex inflationary scenarios, the GW signal could be detectable at scales smaller than the CMB. In the context of the spacetime symmetry breaking, the simple-case effect of explicit diffeomorphism symmetry-breaking terms on PGWs has already been calculated \citep{Nilsson:2022mzq}.
	By computing the modified PGW spectrum and comparing it with standard cosmology, along with sensitivity curves from second and third-generation detectors, we show that BG cosmology improves PGW detection prospects. Additionally, this framework provides sensitivities on BG model parameters, addressing the lack of cosmological constraints.
	
	The manuscript is organized as follows. Section \ref{sec2} provides a concise overview of Bumblebee cosmology in the presence of a timelike vector field. In section \ref{sec3}, we discuss the PGW production in the presence of Bumblebee cosmology. Section \ref{sec4} presents the main results, while Section \ref{sec5} concludes with a summary and discussion of the findings.
	
	We use natural units ($c$ (speed of light in vacuum)$=1= \hbar$ (reduced Planck constant)) and the metric structure $(-,+,+,+)$ throughout the paper unless stated otherwise.

\section{Bumblebee cosmology in presence of a timelike vector field}\label{sec2}
	The Bumblebee model extends GR by incorporating the Bumblebee field $B_{\mu}$, a vector field that interacts non trivially with gravity and acquires a non-zero VEV through a specific potential. This mechanism leads to the SLSB in the gravitational sector, as discussed by Kostelecky \citep{Kostelecky:2003fs}. Taking care of the geometric structures and conservation laws consistent with the standard pseudo-Riemannian manifold admitted in GR, it affects the dynamics of the fields coupled to the Bumblebee field \citep{Kostelecky:2003fs,Bluhm:2004ep}. The simplest model to address LSB in the presence of gravity in a torsion-free
	spacetime is given by the following action in which a single Bumblebee vector field $B_\mu$ coupled to gravity as \citep{Kostelecky:2003fs}
	\begin{equation} \label{action}
    \begin{split}
		S = \int d^{4}x\sqrt{-g} \bigg[ \frac{1}{2\kappa}R + \frac{\xi}{2\kappa}B^{\mu}B^{\nu}R_{\mu\nu} - \frac{1}{4} B_{\mu\nu}B^{\mu\nu} - V(B^{\mu}B_{\mu}\\
        \pm b^{2})+\mathcal{L}_{M} \bigg],
        \end{split}
	\end{equation}
	where $\kappa=8\pi G$, $\xi$ (mass dimension $M^{-2}$) represent the non-minimal coupling constant between Ricci tensor and the Bumblebee field, respectively and $R$ denotes the Ricci scalar. The Bumblebee field strength is defined as $B_{\mu\nu}=\partial_{\mu}B_\nu-\partial_{\nu}B_{\mu}$, analogous to the electromagnetic field. Further, $b^{2} \equiv b^{\mu}b_{\mu} = \langle B^{\mu}B_{\mu}\rangle_{0} 
	\neq 0$ denotes the VEV for the contracted Bumblebee vector field and 
	$V$ is a self-field interaction potential satisfying the condition: $B_{\mu}B^{\mu}\pm b^{2} = 0$, where the sign $\pm$ represent the time-like or space-like nature of $b_{\mu}$ respectively.
	
	The equation of motion of the Bumblebee field is obtained as \citep{Kostelecky:2003fs}
	\begin{equation}\label{BE}
		\nabla_{\mu}B^{\mu \nu} = 2\left(V'B^{\nu}- \frac{\xi}{2\kappa}B_{\mu}R^{\mu\nu}\right)\!,
	\end{equation}
	where the prime over the potential denotes the derivative with respect to its argument i.e., $B_{\mu}$. 
	
	For a time-like Bumblebee vector field $B_{\mu}=\left(B(t)=b,0,0,0\right)$ satisfying relation $B_{\mu} B^{\mu} = \pm\, b^{2}$ \citep{Capelo:2015ipa} in the background of the Bianchi I (BI) metric, the Hubble parameter is obtained as \citep{Sarmah:2024xwx}
	\begin{equation}\label{H}
		H = \sqrt{\frac{\kappa \rho}{{3-\eta(1-2l)}}}\,,
	\end{equation}
	where $\rho= \rho_0 a^{-3 \delta}$ is the energy density with \citep{Sarmah:2024xwx}
	\begin{align}\label{m}
		\delta = &\,\frac{(1+\omega)-l\left\{1+\eta+\omega(1-\eta)\right\}}{(1-l)}-l\,\frac{1 -\frac{\lambda^3}{\alpha \beta} + \frac{18\eta^2(1-2l)}{(1-l)}+9(\gamma - \eta)}{(3+\eta) - l(3+2\eta)} \nonumber\\[5pt]&-\frac{3 l }{2 \left[(3+\eta) - l(3+2\eta)\right]}\left[ 1 + \eta\left(\frac{2l-1}{1-l}\right) \right],
	\end{align}
	where    
    \begin{eqnarray}\label{eta}
\eta &=&\frac{\beta^2-\alpha\beta(1+\beta)+\alpha^2(\beta^2-\beta+1)}{(\alpha+\beta+\alpha\beta)^2}, \nonumber\\ \lambda &=& \frac{\alpha+\beta+\alpha\beta}{3\alpha\beta}\,,  \\
\gamma &=& 3(\alpha\beta\lambda+\eta). \nonumber
	\end{eqnarray}
	Also, $l = \xi b^{2}$ is a free model parameter that characterizes the Lorentz violation, $\eta$ is connected to the shear scalar $\sigma$ associated with the anisotropic spacetime via relation $\sigma^2=\frac{1}{2}[(H_1^2+H_2^2+H_3^2)-3H^2)]$ in which $H_1=\alpha H_2$, $H_1=\beta H_3$, and $H=\frac{H_1+H_2+H_3}{3}$. Here, $\alpha$ and $\beta$ are, in essence, two dimensionless proportionality constants that come from BI metric and sourced the shear scalar $\sigma$. By doing a straightforward calculation, one comes to $\sigma^2=3\eta H^2$. It is clear that $\sigma$, which indicates the anisotropic parameter of spacetime, vanishes when $\eta=0$. This occurs only in the case of $\alpha=1=\beta$, as can be verified by setting $\eta=0$ in (\ref{eta}).
	
	Here, we are interested in preserving the cosmological principle \footnote{The cosmological principle, which is merely a working hypothesis and not a statement representing a fundamental symmetry in physics, assumes the universe is homogeneous and isotropic on large scales, which is captured mathematically by the FLRW metric. Nonetheless, the inhomogeneities anticipated at smaller scales, encompassing galaxies, clusters, voids, and cosmic structures, provide complexities to the gravitational field that are not fully accounted for by the FLRW model.
		This is the origin of the ``fitting problem" presented in the seminal paper \citep{Ellis:1987zz}.  In general,  no fundamental principle guarantees that the metric of the universe is FLRW a priori, meaning that it is always possible to find some anomalies that marginally affect the validity of the cosmological principle reflected in FLRW metric \citep{Krishnan:2021dyb,Krishnan:2021jmh,Secrest:2022uvx,Aluri:2022hzs}. Theoretically, some arguments \citep{Wald:1983ky,Maleknejad:2012as} indicate that if the early universe underwent an inflation phase transition, generally one expects to find a universe close to FLRW, meaning that deviation from the FLRW metric would have challenging outcomes for inflation theory and fundamental physics, as well. In \citep{Green:2014aga}, it is also technically argued that except around strong field objects such as BHs, FLRW usually works well (with an error of only about one part in $10^4$ or even less).}, which signifies the isotropic and homogeneous universe. A hint here is essential. 
	Unlike a space-like Bumblebee vector field as $B_\mu=(0,\overrightarrow{b})$ which violates the homogeneity and isotropy properties of FLRW metric, the presence of the time-like Bumblebee vector field ($B_{\mu} = (b,\overrightarrow{0})$), is not so \citep{Maluf:2021lwh}. It is worth noting that incorporating a space-like Bumblebee vector field into the background of spacetime is phenomenologically interesting, as it introduces anisotropy into the solution and affects the polarization plane of GWs \citep{Amarilo:2023wpn}.
	
	In this regard, by relaxing $\eta$ (i.e., 
	$\alpha=1=\beta$), and using the relation between scale factor $a$, and cosmological redshift $z$ i.e., $a = (1+z)^{-1}$, for the flat FLRW metric, with the following line 
	element
	\begin{equation}\label{}
		ds^2=-dt^2+a(t)^2\bigg(dr^2+r^2d\theta^2+r^2sin^2\theta d\phi^2\bigg),
	\end{equation}
	where the Hubble parameter Eq. \eqref{H} reads off  
	\begin{equation}\label{hub}
		H = H_0 \sqrt{E(z)},
	\end{equation}
	where
	\begin{equation}\label{EZ}
		E(z)=\Omega_{m}(1+z)^{3\delta_m}+\Omega_{r}(1+z)^{3\delta_r}+\Omega_{DE}(1+z)^{3\delta_{DE}},
	\end{equation}
	with
	\begin{align}\label{mr}
		\Bigg\lbrace\delta_m = \frac{2-21l}{2-2l},~~\delta_r &= \frac{8-65l}{6-6l},~~\delta_{DE} = \frac{19l}{2l-2}\Bigg\rbrace.
	\end{align}
	Here, $\Omega_{m,r,DE}$ denote density parameters of matter, radiation, and Dark Energy (DE), respectively. $\Omega_{DE}$, in essence is $1-\Omega_{m}-\Omega_{r}$. 
	The energy density also takes the following form
	\begin{equation}\label{rho}
		\rho=\rho_0 a^{-3(1+\omega)+\frac{57l}{2-2l}}.
	\end{equation}
	As can be seen, in the limit $l=0$, the corresponding GR counterpart of Eqs. (\ref{EZ}), and (\ref{rho}) are recovered as
	\begin{equation}
		E_{GR}(z)=\Omega_{m}(1+z)^{3}+\Omega_{r}(1+z)^{4}+\Omega_{DE},~~~~~\rho=\rho_0 a^{-3(1+\omega)} \,,
		\label{E_s}
	\end{equation}
	where we use Eqs. \eqref{hub}, \eqref{EZ}, and \eqref{mr} for studying PGW in presence of time-like Bumblebee vector field \footnote{In general, a $\omega$CDM model is considered in which the universe's energy density budget consists of three parts: $\rho = \rho_{m} + \rho_{r}+\rho_{DE}$, where the contribution of each of them come from $\rho =\rho_0 a^{-3\delta}$.
		The general form of $\delta$ is given by Eq. \eqref{m}.  Particularly, the contribution of DE density of energy is given by  $\rho_{DE} =\rho_0 a^{\delta_{DE}}$ in which $\delta_{DE}$ is obtain by setting $\alpha=1=\beta$ or $\eta=0$ and $\omega=-1$ into Eq. \eqref{m} i.e., $\delta_{DE}=\frac{19l}{2l-2}$ (as written in Eq. \eqref{mr}.}.	
	
	A hint on the range of the model parameters $\delta_{m,r,DE}$, and $l$ is essential. The impact of SLSB is treated as a minor perturbation in the universe's cosmological history, leading to slight deviations from the parameters of Standard Cosmology (SC). Therefore, values of $\delta_{m,r,DE}$ in Eq. \eqref{EZ} must be around $1, 4/3, 0$, respectively. In this way, we limit these parameters within ranges: $0.9<\delta_{m}<1.05$, $1.25<\delta_{r}<1.75$, and $-10^{-3}<\delta_{DE}<10^{-2}$. By taking these constraints into account of Eq. (\ref{mr}), we find that the constraint obtained on $l$ has just a slight sensitivity to range $\delta_{DE}$.	For values $\delta_{DE}\geq10^{-2}$, the allowed range of BG parameter lies within $-5\times 10^{-3}\lesssim l\lesssim 10^{-4}$, while, for case of $\delta_{DE}<10^{-2}$, it slightly changes to $-10^{-3}\leq l\leq 10^{-4}$. This connection between $l$, and $\delta_{DE}$, although slight, may potentially signal us about the understanding of the phase of accelerated expansion of the universe via an SLSB gravitational model \citep{Jesus:2019nwi}. 
	
	In general, there is still no agreement on the sign of $l$, and both positive and negative values are recommended, depending on the phenomenological framework under consideration. Some previous studies have shown that BG supported by $l<0$, potentially is a rich framework for justifying some astrophysical phenomena (e.g., see Refs. \citep{Khodadi:2021owg,Khodadi:2022dff,Khodadi:2023yiw}). In the present work, we will show that this is also the case for the detection of PGWs. However, $l<0$ may contain some worrying, albeit justifiable, consequences. Authors of \citep{Delhom:2019wcm} in the framework of a metric-affine formulation of BG, have evaluated the phenomenological effects arising from the coupling between the Bumblebee vector field and scalar and Dirac fields. For the scalar fields, if the Bumblebee vector field is time-like, then $l<0$, results in the appearance of ghost/tachyon-like instabilities. At first sight, it may seem worrying, but by incorporating the higher-order terms in the calculation, one may get rid of this problem, as discussed in \citep{Delhom:2019wcm}. For the Dirac fields case, the negative value of $l$ does not make trouble within the range $-1\leq l<0$. This does not mean that the $l>0$ is immune to pathological behavior. An instance in \citep{Mai:2024lgk} demonstrated that a Bumblebee BH supporting $l>0$ with the Bumblebee charge may, under some conditions, be unstable, particularly gradient and tachyonic instabilities.

	\section{Primordial gravitational waves in Bumblebee cosmology}\label{sec3}
	The quantity that parametrizes the inflation-driven PGW is the power spectrum of primordial tensor perturbations generated during inflation. In this section, we will evaluate the contribution of SLSB arising from a time-like Bumblebee vector field on the strain amplitude of PGW signal. Bumblebee cosmology is indeed an early universe scenario based on modified gravity that includes the Bumblebee vector field (here time-like) in the background of spacetime. The main point is that modifications of GR naturally imply variations in the Hubble expansion rate of the universe at the metric level of the theory, leaving an imprint on any processes that convey information about the expansion history and potentially produce observational signatures. Furthermore, the propagation of PGW transmits information about the universe's expansion history through its evolution in the post-inflationary phase. Thus, we investigated this probe of the early universe expansion rate to understand the PGW signals predicted in the Bumblebee modified gravity theory of the early universe.

	Let us assume that the tensor perturbations $h_{00} =  h_{0i} = 0$. Therefore, we will be working in the Transverse Traceless ($TT$) gauge, where $\partial^i h_{ij}=0$ and $h^i_i=0$. Hence, the dynamics of the tensor perturbation in first-order perturbation theory i.e., the linearized equation is then obtained as \citep{Watanabe:2006qe} (see also seminal paper \citep{Mukhanov:1990me})
	\begin{equation}
	\label{hdyn}
	\ddot h_{ij}\ + \ 3H\dot h_{ij} \ - \ \frac{\nabla^2}{a^2}h_{ij}\ = \ 16\pi G\hspace{0.3mm} \Pi_{ij}^{TT}\,,
	\end{equation}
	with $\Pi_{ij}^{TT}$ is the $TT$ anisotropic part of the stress tensor, given as
	\begin{equation}
	\Pi_{ij} \ = \ \frac{T_{ij}-p\hspace{0.3mm} g_{ij}}{a^2}\,,
	\label{TPE}
	\end{equation}
	where $T_{ij}$, $g_{ij}$ and $p$ are the stress-energy tensor, the metric tensor and the background pressure, respectively. The expression Eq. (\ref{hdyn}) is valid as long as $\Pi_{ij}^{TT}$ acts as a perturbation around a perfect fluid \citep{Dodelson:2003ft}. 
	
	To solve Eq. \eqref{hdyn}, we work in the Fourier space, 
	where \citep{Watanabe:2006qe}
	\begin{equation}
	h_{ij}(t,\vec{x}) \ = \ \sum_{\lambda}\int\frac{d^3k}{\left(2\pi\right)^3}\hspace{0.2mm}h^\lambda(t,\vec{k})\hspace{0.2mm}\epsilon^\lambda_{ij}(\vec{k})\hspace{0.2mm}e^{i\vec{k}\cdot\vec{x}}\,,
	\end{equation}
	where $\epsilon^\lambda$ is the spin-2
	polarization tensor obeying the orthonormality condition $\sum_{ij}\epsilon^\lambda_{ij}\epsilon^{\lambda'*}_{ij}=2\delta^{\lambda\lambda'}$ and $\lambda=+,\times$ is the two independent wave-polarizations.
	
	The tensor perturbation $h^\lambda(t,\vec{k})$ can be written as
	\begin{equation}
	h^\lambda(t,\vec{k}) = h_{\mathrm{prim}}^\lambda(\vec{k})X(t,k)\,,
	\end{equation}
	where $k = |\vec{k}|$ represents the magnitude of the wave vector, $X(t,k)$ is the transfer function that describes the time-dependent evolution of the perturbation, and $h_{\mathrm{prim}}^\lambda$ denotes the amplitude of the primordial tensor perturbations. Using this parametrization, the tensor power spectrum can be expressed as~\citep{Bernal:2019lpc,Bernal:2020ywq}
	\begin{equation}
	\mathcal{P}_T(k) \ = \ \frac{k^3}{\pi^2}\sum_\lambda\Big|h^\lambda_{\mathrm{prim}}(\vec k)\Big|^2 \ = \ \frac{2}{\pi^2}\hspace{0.3mm}G\hspace{0.3mm} H^2\Big|_{k=aH}\,.
	\end{equation}
	As a result, Eq.~\eqref{hdyn} can be rewritten in the form of a damped harmonic oscillator equation as
	\begin{equation}\label{XGae}
	X'' \ + \ 2\hspace{0.2mm}\frac{a'}{a}X' \ + \ k^2X \ = \ 0\,,
	\end{equation}
	where the prime denotes differentiation with respect to the conformal time $\tau$, defined by the relation $d\tau = dt/a$. The right-hand side of Eq. \eqref{hdyn} has been set to zero as it is not relevant for our studies in the chosen frequency range. In a flat Universe, where the fluid dominates the matter content, the scale factor evolves as $a(\tau)\propto \tau^{\frac{2}{1+3w}}$. This allows the damping term to be expressed as
	\begin{equation}
	\label{damp}
	2\hspace{0.2mm}\frac{a'}{a} \ = \ \frac{4}{\tau\left(1+3w\right)}\,,
	\end{equation}
	where $w$ represents the Equation of State (EoS) parameter of the fluid.
	
	Hence, the relic density of PGW from first-order tensor perturbation in the SC reads \citep{Bernal:2020ywq,Watanabe:2006qe}
	\begin{eqnarray}
		\mbox{\hspace{-7mm}}\Omega_{\mathrm{GW}}
		(\tau,k)=\frac{[X'(\tau,k)]^2}{12a^2(\tau)H^2(\tau)}\,\mathcal{P}_T(k)
		\simeq\left[\frac{a_{\mathrm{hc}}}{a(\tau)}\right]^4\left[\frac{H_{\mathrm{hc}}}{H(\tau)}\right]^2\frac{\mathcal{P}_T(k)}{24},
		\label{Ttps}
	\end{eqnarray}
	where we average over periods of oscillations, implies
	\begin{equation}\label{eq:x}
	X'(\tau,k)\ \simeq \ k\hspace{0.2mm} X(\tau,k)\ \simeq \  \frac{k\hspace{0.3mm} a_{\mathrm{hc}}}{\sqrt{2}a(\tau)}\ \simeq \
	\frac{a^2_{\mathrm{hc}}\hspace{0.3mm}H_{\mathrm{hc}}}{\sqrt{2}a(\tau)}\,,
	\end{equation}
	with $k=2\pi f=a_{\mathrm{hc}}H_{\mathrm{hc}}$ at the horizon crossing. Therefore, the PGW relic density at the present time is obtained as 
	\begin{eqnarray}
		\mbox{\hspace{-7mm}}\Omega_{\mathrm{GW}}(\tau_0,k)h^2\simeq\left[\frac{g_*(T_{\mathrm{hc}})}{2}\right]\left[\frac{g_{*s}(T_0)}{g_{*s}(T_{\mathrm{hc}})}\right]^{4/3}\frac{\mathcal{P}_T(k)\Omega_{r}(T_0)h^2}{24}\,,
		\label{Spt}
	\end{eqnarray}
	where $h$ is the dimensionless Hubble constant, $\Omega_r=\rho_r/\rho_{cr}$ with $\rho_{cr}=3H^2_0/8\pi G$ denotes the critical radiation energy density, and $g_*(T)$ and $g_{*s}(T)$ the effective numbers of relativistic degrees of freedom that contribute to the radiation energy density $\rho_r$ and entropy density $s_r$, respectively, given as,
	\begin{eqnarray}
		\rho_r=\frac{\pi^2}{30}g_*(T)T^4\,,~s_r=\frac{2\pi^2}{45}g_{*s}(T)T^3\,.
	\end{eqnarray}
	
	The scale dependence of the tensor power spectrum is defined as
	\begin{equation}
	\mathcal{P}_T(k) \ = \ A_T\left(\frac{k}{\tilde k}\right)^{n_T}\,,
	\end{equation}
	where $n_T$ is the tensor spectral index \footnote{While $n_T=0$ indicates a scale-invariant primordial tensor spectrum, the cases of $n_T>0$, and $n_T<0$, respectively describe blue-tilt spectrum (amplified PGWs), and a red-tilt spectrum (suppressed PGWs) \citep{Vagnozzi:2023lwo}.} and $\tilde k=0.05\,\mathrm{Mpc}^{-1}$ is a characteristic wave number scale or the Planck 2018 pivot scale. The
	amplitude of the tensor perturbation $A_T$ is related to the scalar perturbation amplitude $A_S$ by $A_T=r A_S$, with $r$ being the tensor-to-scalar ratio. 
	
	To find the role of SLSB originating from a time-like Bumblebee vector field in the FRW background on the PGW spectrum, we use Eq. \eqref{hub} in Eq.~\eqref{Ttps} as
	\begin{eqnarray}
		\Omega_{\mathrm{GW}}(\tau,k)\!&\simeq& \!\left[\frac{a_{\mathrm{hc}}}{a(\tau)}\right]^4\left[\frac{H_{\mathrm{hc}}}{H_{\mathrm{GR}}(\tau)}\right]^2\left[\frac{H_{\mathrm{GR}}(\tau)}{H(\tau)}\right]^2\frac{\mathcal{P}_T(k)}{24} \nonumber \\[2mm]
		&=&\!\Omega^{\mathrm{GR}}_{\mathrm{GW}}(\tau,k)\left[\frac{H_{\mathrm{GR}}(\tau)}{H(\tau)}\right]^2\left[ \frac{a_{\mathrm{hc}}}{a_{\mathrm{hc}}^{\mathrm{GR}}}\right]^4\ \ \left[ \frac{a^{\mathrm{GR}}(\tau)}{a(\tau)}\right]^4\times \nonumber\\
       \left[ \frac{H_{\mathrm{hc}}}{H_{\mathrm{hc}}^{\mathrm{GR}}}\right]^2   \,\hspace{-6.5 cm},
		\label{eq:PGWBar0}
	\end{eqnarray}
	where ``GR" in the subscript/superscript refers to quantities defined within the standard GR framework. For example, $\Omega^{\mathrm{GR}}_{\mathrm{GW}}(\tau, k)$ represents the predicted relic density of PGWs according to conventional GR, which matches Eq. (\ref{Ttps}). Given that $E(z) = 1$ at $z = 0$, as shown in Eq.~\eqref{EZ} and conformal time $\tau_0$, we can express the GW energy density as
	\begin{equation}
		\Omega_{\mathrm{GW}}(\tau_0,k)
		\ \simeq \ \Omega^{\mathrm{GR}}_{\mathrm{GW}}(\tau_0,k)\left[ \frac{a_{\mathrm{hc}}}{a_{\mathrm{hc}}^{\mathrm{GR}}}\right]^4\left[ \frac{H_{\mathrm{hc}}}{H_{\mathrm{hc}}^{\mathrm{GR}}}\right]^2 \,. \label{eq:PGWBar} 
	\end{equation}
	\section{Sensitivities of Lorentz-violating parameter in the search for primordial gravitational waves} \label{sec4}
	In this section, we focus on GWs within the frequency range of $(10^{-10}~\mathrm{Hz},10^4~\mathrm{Hz})$. We exclude contributions from frequencies lower than $10^{-10}~\mathrm{Hz}$, which arise from the free streaming of neutrinos and photons~\citep{Weinberg:2003ur}\footnote{In Ref. \citep{Weinberg:2003ur}, it is shown that a cosmological background with an anisotropic stress tensor, suitable for a free-streaming thermal massless neutrino background, can damp PGWs after they enter the horizon. In Ref. \citep{Dent:2013asa}, it has been generalized to the case of the massive neutrino, extra neutrino species, and a possible relativistic background of axions. }. The chosen frequency range in our case is anticipated to be thoroughly explored by current and forthcoming GW observatories including NANOGrav \citep{NANOGrav:2023gor}, Square Kilometer Array (SKA) \citep{Janssen:2014dka}, THEIA \citep{Theia:2019non}, $\mu$-Ares \citep{Sesana:2019vho}, ASTRO-GW \citep{Ni:2012eh}, Atom Interferometer Observatory and Network (AION-Km) \citep{Badurina:2019hst}, LISA interferometer~\citep{LISA:2017pwj},  DECi-hertz Interferometer Gravitational wave Observatory (DECIGO) \citep{Kawamura:2020pcg}, Atomic Experiment for Dark Matter and Gravity Exploration (AEDGE) \citep{AEDGE:2019nxb}, Big Bang Observer (BBO) \citep{Crowder:2005nr}, Einstein Telescope (ET) \citep{Sathyaprakash:2012jk}, Cosmic Explorer (CE) \citep{Evans:2021gyd}, and Advanced LIGO (aLIGO) \citep{LIGOScientific:2014pky}.
	
	\begin{figure}
		\begin{center}			\hspace{-1mm}\includegraphics[width=90 mm]{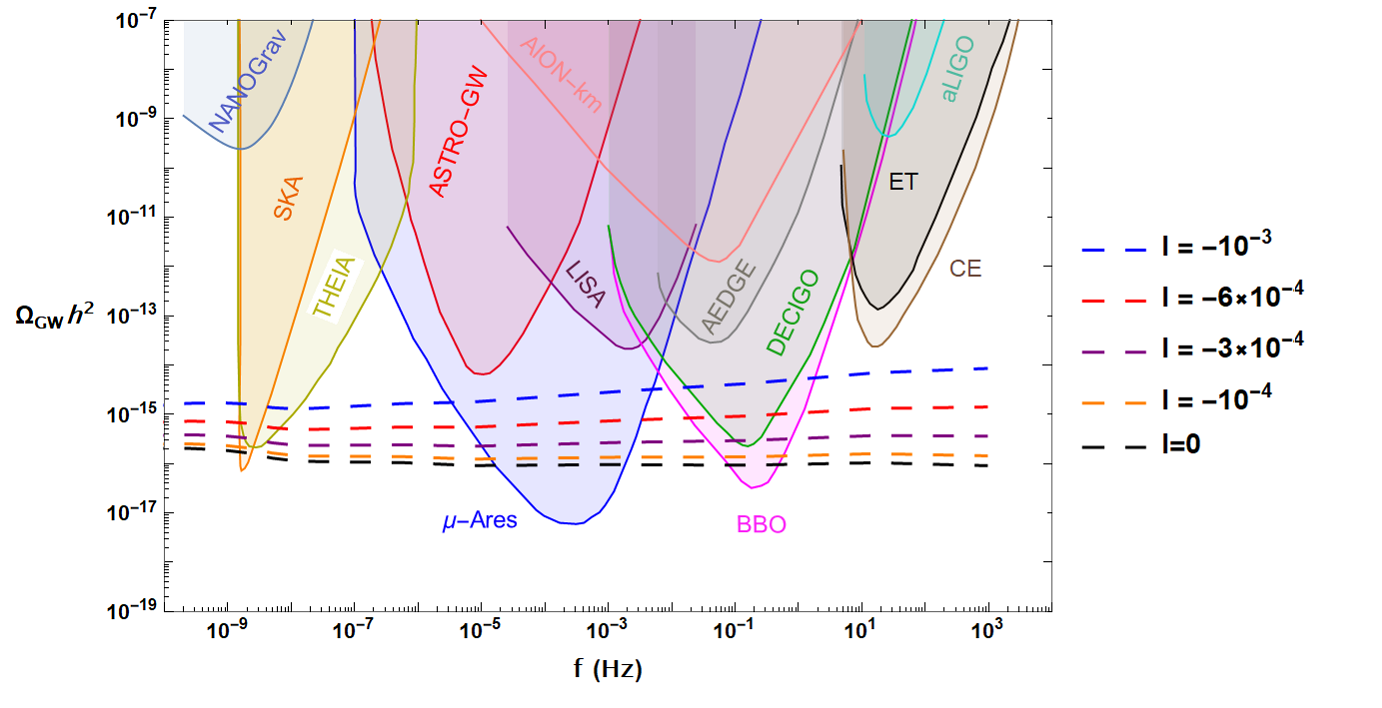}~~~
			\caption{Plot of the strain amplitude of the PGW spectrum as a function of frequency $f$ for the Bumblebee cosmology with different values of $l$ within range $-10^{-3}\leq l\leq 0$. The colored shaded areas denote different sensitivity regions for upcoming GW detectors. }
			\label{Fig1}
		\end{center}
	\end{figure}
	
	\begin{figure}
		\begin{center}
	\hspace{-1mm}\includegraphics[width=90 mm]{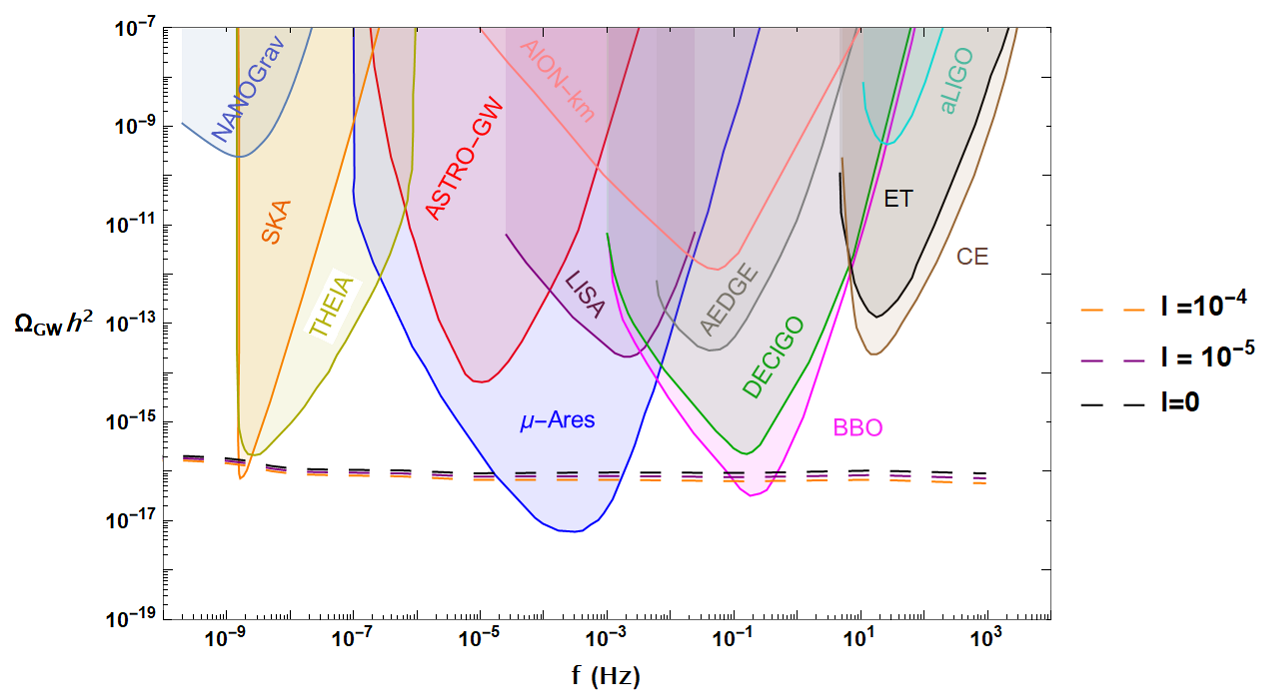}
			\caption{Plot of the strain amplitude of the PGW spectrum as a function of frequency $f$ for the Bumblebee cosmology with different values of $l$ within the range $0\leq l\leq 10^{-4}$. The colored shaded areas denote different sensitivity regions for upcoming GW detectors.}
			\label{Fig2}
		\end{center}
	\end{figure}

	Using Eq. \eqref{eq:PGWBar0}, and taking the values of $l$ in the allowed ranges $-10^{-3}\leq l\leq0$ (Figure \ref{Fig1}) and $0\leq l\leq 10^{-4}$ ( Figure \ref{Fig2}), we show the PGW spectrum, $\Omega_{\mathrm{GW}}h^2$ as a function of frequency $f$ for various values of $l$ within the aforementioned allowed frequency ranges. The plot is obtained using Eq.~\eqref{eq:PGWBar}, where one has to relate the frequency to the $z$ value with the equation $k =H(z)/(z+ 1)$. Inverting the latter numerically, it is possible to obtain $H_{hc}$, $H_{hc}^{GR}$, $a_{hc}$ and $a_{hc}^{GR}$ as functions of $k$. Operatively, we construct one vector with the $z$ and $k$ values. These vectors are used to obtain a function $k(z)$. Once done, we can construct another vector with $k(z)$ and $\Omega_{\rm GW}(\tau_0,k(z))/\Omega_{\rm GW}^{\rm GR}(\tau_0,k(z))$ values, as expressed in Eq.~\eqref{eq:PGWBar}. These values are plotted in Figures~\ref{Fig1} and \ref{Fig2}. To draw these plots we have used numerical values: $ H_0 = 100 \times h_0 ~ =67.4~\rm km (s~Mpc)^{-1}, \Omega_m = 0.044,~ \Omega_r = 9\times 10^{-5}, ~  T_0 = 2.725 K,~ M_p = 1.22\times 10^{19}~GeV$. A note is necessary here. The relationship between the frequency $f$ and redshift $z$ is given by the equation immediately below Eq. \eqref{eq:x}. Indeed, the equation $2\pi f=a_{\mathrm{hc}}H_{\mathrm{hc}}$ directly links the frequency to $a$ and $H$, both of which can be expressed as functions of $z$. In this way, it is possible to derive a relation between $f$ and $z$, which cannot be expressed analytically but can be evaluated numerically.
	
	It is evident that the possibility of probing the Bumblebee cosmology-based PGW by the upcoming detectors with different frequency sensitivities depends on the values of $l$. The strain amplitude of the PGW spectrum arising from a Bumblebee vector field for negative values of $l$ within the range $(-10^{-3}\leq l\leq 0)$ is bigger than its counterpart in SC. The dashed-black line in Figure \ref{Fig1} (also Figure \ref{Fig2}) displays the spectrum Eq. \eqref{Spt} as a function of frequency $f$, assuming a scale-invariant primordial tensor spectrum ($n_T = 0$) and consistent with the Planck observational constraint $A_S \simeq 2.1 \times 10^{-9}$ at the CMB scale \citep{Planck:2018gnk}. For positive values of $l$, the Bumblebee vector field does not make any constructive contribution in terms of detection to the standard PGW spectrum as the amplitude is suppressed slightly.
	
	As it is evident from Figure \ref{Fig1}, the practical distinguishability of the modified PGW signals from the standard ones is promising, as a negative \( l \) (\( l < 0 \)) can enhance the strain amplitude by an order of magnitude. In this scenario, where the standard strain amplitude is enhanced, experiments like SKA, $\mu$-Ares, and BBO are anticipated to have an improved likelihood of detecting PGW signals. Furthermore, the modified PGW signals may also intersect the sensitivity curves of experiments such as THEIA and DECIGO, in addition to the aforementioned detectors, potentially differentiating with $l>0$ scenario.
	
	\section{Conclusions and discussions}\label{sec5}
	This manuscript investigates how LIV, modeled via Bumblebee cosmology with a time-like vector field acquiring a nonzero VEV, modifies the PGW spectrum from inflation and affects its detectability. We explore the sensitivity of the dimensionless Lorentz violation parameter (\(l\)) across a range of current and upcoming detectors, including NANOGrav, SKA, THEIA, \(\mu\)-Ares, ASTRO-GW, AION-Km, LISA, DECIGO, AEDGE, BBO, ET, CE, and aLIGO. Our analysis spans the frequency range of \((10^{-10}~\mathrm{Hz}, 10^4~\mathrm{Hz})\), providing insights into the potential for probing LIV with these advanced instruments.
	
	We analyze \(l\) within the range \(-10^{-3} \leq l \leq 10^{-4}\), encompassing both positive and negative values, and investigate the detection prospects of PGWs using current and future GW detectors. For \(-10^{-3} \leq l \leq 0\) (negative \(l\)), the PGW spectrum exhibits a significant amplification. Notably, at \(l \sim -10^{-3}\), the strain amplitude of PGWs increases by an order of magnitude compared to the \(l=0\) case, which corresponds to SC. In contrast, for positive values of \(l\) (\(0 \leq l \leq 10^{-4}\)), the PGW spectrum is suppressed compared to the \(l=0\) scenario. 

    As the cosmic string tension increases, the energy density of the PGW spectrum sourced by cosmic strings also rises. However, the spectral shape and tilt provide key signatures to distinguish PGWs from different origins. For instance, the PGW spectrum from cosmic strings exhibits multiple kinks due to varying loop sizes and reconnections, along with a strong frequency dependence
    \citep{Battye1998,Chang:2021afa,Auclair:2022ylu,Yonemaru:2020bmr,LIGOScientific:2021nrg,Meijer:2023yhn}. In contrast, the PGW spectrum generated in Lorentz-violating Bumblebee gravity lacks these features, presenting a smoother profile.
    
	Both positive and negative values of \(l\) are within the sensitivity ranges of detectors such as SKA, \(\mu\)-Ares, and BBO. However, THEIA and DECIGO are exclusively sensitive to negative \(l\) values, implying that any signal detected by these instruments could provide evidence for Lorentz violation driven by a time-like Bumblebee vector field with negative \(l\).

The predicted values of $ \Omega_{GW}$ are constrained by the frequency ranges accessible to specific experimental technologies. Moreover, the theoretical PGW spectrum in Bumblebee cosmology--where the strain amplitude $ \Omega_{GW}$ is modified by the LSB parameter--is derived from Eq. (\ref{eq:PGWBar0}). By placing the theoretical spectrum in the experimental frequency range, it can be seen whether the modification appearing in the PGW improves the chances of detecting the standard PGW. 
	As usual, we just do a numerical comparison between the $ \Omega_{GW}$ of PGW produced by the underlying theory with sensitivity curves expected from experiments. It has been shown that not all underlying experiments can detect the modified PGW spectrum or the standard sample. In other words, if any of these PGW spectra with the given value of $ \Omega_{GW}$ are detected by one of the allowed detectors in a relevant frequency range, it means that it potentially comes from Bumblebee cosmology at hand or standard cosmology. In this way, one can discriminate LSB from LS conservation. Concerning the case of $l>0$, the deviation of LSB from LS conservation (standard cosmology) by the detectors is very negligible and hardly separable.	An important point to note is that there is no guarantee that these spectra will be detected simultaneously in all potentially allowed experiments because the frequency range of each is different. 
	It is necessary to emphasize that the similar values of $\Omega_{GW}$ may be produced by any potential modifications in the early universe. As a result, we assert that if the underlying Bumblebee cosmology is correct, then the LSB parameter embedded within it can produce certain PGW spectra detectable by some experiments. Its inverse is incorrect, meaning that the detection of a spectrum modified by the LSB parameter does not demonstrate that Bumblebee cosmology is the correct theory according to the experiments.

\section*{Acknowledgements}
The authors would like to thank Pranjal Sarmah for the fruitful discussions. The authors also wish to thank the referees for their constructive comments on improving the quality of this manuscript. G.L. and T.K.P. thank COST Actions CA21106 (COSMIC WISPers), and CA23130 (BridgeQG) supported by COST (European Cooperation in Science and Technology).




\bibliographystyle{elsarticle-harv} 
\bibliography{example}

\begin{thebibliography}{136}
\expandafter\ifx\csname natexlab\endcsname\relax\def\natexlab#1{#1}\fi
\providecommand{\url}[1]{\texttt{#1}}
\providecommand{\href}[2]{#2}
\providecommand{\path}[1]{#1}
\providecommand{\DOIprefix}{doi:}
\providecommand{\ArXivprefix}{arXiv:}
\providecommand{\URLprefix}{URL: }
\providecommand{\Pubmedprefix}{pmid:}
\providecommand{\doi}[1]{\href{http://dx.doi.org/#1}{\path{#1}}}
\providecommand{\Pubmed}[1]{\href{pmid:#1}{\path{#1}}}
\providecommand{\bibinfo}[2]{#2}
\ifx\xfnm\relax \def\xfnm[#1]{\unskip,\space#1}\fi
\bibitem[{Aasi et~al.(2015)}]{LIGOScientific:2014pky}
\bibinfo{author}{Aasi, J.}, et~al. (\bibinfo{collaboration}{LIGO Scientific}),
  \bibinfo{year}{2015}.
\newblock \bibinfo{title}{{Advanced LIGO}}.
\newblock \bibinfo{journal}{Class. Quant. Grav.} \bibinfo{volume}{32},
  \bibinfo{pages}{074001}.
\newblock \DOIprefix\doi{10.1088/0264-9381/32/7/074001},
  \href{http://arxiv.org/abs/1411.4547}{{\tt arXiv:1411.4547}}.
\bibitem[{Abbott et~al.(2019)}]{LIGOScientific:2019vic}
\bibinfo{author}{Abbott, B.P.}, et~al. (\bibinfo{collaboration}{LIGO
  Scientific, Virgo}), \bibinfo{year}{2019}.
\newblock \bibinfo{title}{{Search for the isotropic stochastic background using
  data from Advanced LIGO\textquoteright{}s second observing run}}.
\newblock \bibinfo{journal}{Phys. Rev. D} \bibinfo{volume}{100},
  \bibinfo{pages}{061101}.
\newblock \DOIprefix\doi{10.1103/PhysRevD.100.061101},
  \href{http://arxiv.org/abs/1903.02886}{{\tt arXiv:1903.02886}}.
\bibitem[{Abbott et~al.(2021)}]{LIGOScientific:2021nrg}
\bibinfo{author}{Abbott, R.}, et~al. (\bibinfo{collaboration}{LIGO Scientific,
  Virgo, KAGRA}), \bibinfo{year}{2021}.
\newblock \bibinfo{title}{{Constraints on Cosmic Strings Using Data from the
  Third Advanced LIGO\textendash{}Virgo Observing Run}}.
\newblock \bibinfo{journal}{Phys. Rev. Lett.} \bibinfo{volume}{126},
  \bibinfo{pages}{241102}.
\newblock \DOIprefix\doi{10.1103/PhysRevLett.126.241102},
  \href{http://arxiv.org/abs/2101.12248}{{\tt arXiv:2101.12248}}.
\bibitem[{Ach\'ucarro et~al.(2022)}]{Achucarro:2022qrl}
\bibinfo{author}{Ach\'ucarro, A.}, et~al., \bibinfo{year}{2022}.
\newblock \bibinfo{title}{{Inflation: Theory and Observations}}
  \href{http://arxiv.org/abs/2203.08128}{{\tt arXiv:2203.08128}}.
\bibitem[{Agazie et~al.(2023)}]{NANOGrav:2023gor}
\bibinfo{author}{Agazie, G.}, et~al. (\bibinfo{collaboration}{NANOGrav}),
  \bibinfo{year}{2023}.
\newblock \bibinfo{title}{{The NANOGrav 15 yr Data Set: Evidence for a
  Gravitational-wave Background}}.
\newblock \bibinfo{journal}{Astrophys. J. Lett.} \bibinfo{volume}{951},
  \bibinfo{pages}{L8}.
\newblock \DOIprefix\doi{10.3847/2041-8213/acdac6},
  \href{http://arxiv.org/abs/2306.16213}{{\tt arXiv:2306.16213}}.
\bibitem[{Ahmadvand and Bitaghsir~Fadafan(2017)}]{Ahmadvand:2017xrw}
\bibinfo{author}{Ahmadvand, M.}, \bibinfo{author}{Bitaghsir~Fadafan, K.},
  \bibinfo{year}{2017}.
\newblock \bibinfo{title}{{Gravitational waves generated from the cosmological
  QCD phase transition within AdS/QCD}}.
\newblock \bibinfo{journal}{Phys. Lett. B} \bibinfo{volume}{772},
  \bibinfo{pages}{747--751}.
\newblock \DOIprefix\doi{10.1016/j.physletb.2017.07.039},
  \href{http://arxiv.org/abs/1703.02801}{{\tt arXiv:1703.02801}}.
\bibitem[{Ahmadvand and Bitaghsir~Fadafan(2018)}]{Ahmadvand:2017tue}
\bibinfo{author}{Ahmadvand, M.}, \bibinfo{author}{Bitaghsir~Fadafan, K.},
  \bibinfo{year}{2018}.
\newblock \bibinfo{title}{{The cosmic QCD phase transition with dense matter
  and its gravitational waves from holography}}.
\newblock \bibinfo{journal}{Phys. Lett. B} \bibinfo{volume}{779},
  \bibinfo{pages}{1--8}.
\newblock \DOIprefix\doi{10.1016/j.physletb.2018.01.066},
  \href{http://arxiv.org/abs/1707.05068}{{\tt arXiv:1707.05068}}.
\bibitem[{Akrami et~al.(2020)}]{Planck:2018gnk}
\bibinfo{author}{Akrami, Y.}, et~al. (\bibinfo{collaboration}{Planck}),
  \bibinfo{year}{2020}.
\newblock \bibinfo{title}{{Planck 2018 results. XI. Polarized dust
  foregrounds}}.
\newblock \bibinfo{journal}{Astron. Astrophys.} \bibinfo{volume}{641},
  \bibinfo{pages}{A11}.
\newblock \DOIprefix\doi{10.1051/0004-6361/201832618},
  \href{http://arxiv.org/abs/1801.04945}{{\tt arXiv:1801.04945}}.
\bibitem[{Aluri et~al.(2023)}]{Aluri:2022hzs}
\bibinfo{author}{Aluri, P.K.}, et~al., \bibinfo{year}{2023}.
\newblock \bibinfo{title}{{Is the observable Universe consistent with the
  cosmological principle?}}
\newblock \bibinfo{journal}{Class. Quant. Grav.} \bibinfo{volume}{40},
  \bibinfo{pages}{094001}.
\newblock \DOIprefix\doi{10.1088/1361-6382/acbefc},
  \href{http://arxiv.org/abs/2207.05765}{{\tt arXiv:2207.05765}}.
\bibitem[{Amarilo et~al.(2024)Amarilo, Filho, Filho and Reis}]{Amarilo:2023wpn}
\bibinfo{author}{Amarilo, K.M.}, \bibinfo{author}{Filho, M.B.F.},
  \bibinfo{author}{Filho, A.A.A.}, \bibinfo{author}{Reis, J.A.A.S.},
  \bibinfo{year}{2024}.
\newblock \bibinfo{title}{{Gravitational waves effects in a
  Lorentz\textendash{}violating scenario}}.
\newblock \bibinfo{journal}{Phys. Lett. B} \bibinfo{volume}{855},
  \bibinfo{pages}{138785}.
\newblock \DOIprefix\doi{10.1016/j.physletb.2024.138785},
  \href{http://arxiv.org/abs/2307.10937}{{\tt arXiv:2307.10937}}.
\bibitem[{Amaro-Seoane et~al.(2017)}]{LISA:2017pwj}
\bibinfo{author}{Amaro-Seoane, P.}, et~al. (\bibinfo{collaboration}{LISA}),
  \bibinfo{year}{2017}.
\newblock \bibinfo{title}{{Laser Interferometer Space Antenna}}
  \href{http://arxiv.org/abs/1702.00786}{{\tt arXiv:1702.00786}}.
\bibitem[{Anand et~al.(2017)Anand, Dey and Mohanty}]{Anand:2017kar}
\bibinfo{author}{Anand, S.}, \bibinfo{author}{Dey, U.K.},
  \bibinfo{author}{Mohanty, S.}, \bibinfo{year}{2017}.
\newblock \bibinfo{title}{{Effects of QCD Equation of State on the Stochastic
  Gravitational Wave Background}}.
\newblock \bibinfo{journal}{JCAP} \bibinfo{volume}{03}, \bibinfo{pages}{018}.
\newblock \DOIprefix\doi{10.1088/1475-7516/2017/03/018},
  \href{http://arxiv.org/abs/1701.02300}{{\tt arXiv:1701.02300}}.
\bibitem[{Antoniadis et~al.(2024)}]{EPTA:2023xxk}
\bibinfo{author}{Antoniadis, J.}, et~al. (\bibinfo{collaboration}{EPTA,
  InPTA}), \bibinfo{year}{2024}.
\newblock \bibinfo{title}{{The second data release from the European Pulsar
  Timing Array - IV. Implications for massive black holes, dark matter, and the
  early Universe}}.
\newblock \bibinfo{journal}{Astron. Astrophys.} \bibinfo{volume}{685},
  \bibinfo{pages}{A94}.
\newblock \DOIprefix\doi{10.1051/0004-6361/202347433},
  \href{http://arxiv.org/abs/2306.16227}{{\tt arXiv:2306.16227}}.
\bibitem[{Aoki et~al.(2017)Aoki, Goto and Kubo}]{Aoki:2017aws}
\bibinfo{author}{Aoki, M.}, \bibinfo{author}{Goto, H.}, \bibinfo{author}{Kubo,
  J.}, \bibinfo{year}{2017}.
\newblock \bibinfo{title}{{Gravitational Waves from Hidden QCD Phase
  Transition}}.
\newblock \bibinfo{journal}{Phys. Rev. D} \bibinfo{volume}{96},
  \bibinfo{pages}{075045}.
\newblock \DOIprefix\doi{10.1103/PhysRevD.96.075045},
  \href{http://arxiv.org/abs/1709.07572}{{\tt arXiv:1709.07572}}.
\bibitem[{Apreda et~al.(2002)Apreda, Maggiore, Nicolis and
  Riotto}]{Apreda:2001us}
\bibinfo{author}{Apreda, R.}, \bibinfo{author}{Maggiore, M.},
  \bibinfo{author}{Nicolis, A.}, \bibinfo{author}{Riotto, A.},
  \bibinfo{year}{2002}.
\newblock \bibinfo{title}{{Gravitational waves from electroweak phase
  transitions}}.
\newblock \bibinfo{journal}{Nucl. Phys. B} \bibinfo{volume}{631},
  \bibinfo{pages}{342--368}.
\newblock \DOIprefix\doi{10.1016/S0550-3213(02)00264-X},
  \href{http://arxiv.org/abs/gr-qc/0107033}{{\tt arXiv:gr-qc/0107033}}.
\bibitem[{Ara\'ujo~Filho(2025)}]{AraujoFilho:2025hkm}
\bibinfo{author}{Ara\'ujo~Filho, A.A.}, \bibinfo{year}{2025}.
\newblock \bibinfo{title}{{How does non-metricity affect particle creation and
  evaporation in bumblebee gravity?}}
  \href{http://arxiv.org/abs/2501.00927}{{\tt arXiv:2501.00927}}.
\bibitem[{Ara\'ujo~Filho et~al.(2024)Ara\'ujo~Filho, Nascimento, Petrov and
  Porf\'\i{}rio}]{AraujoFilho:2024ykw}
\bibinfo{author}{Ara\'ujo~Filho, A.A.}, \bibinfo{author}{Nascimento, J.R.},
  \bibinfo{author}{Petrov, A.Y.}, \bibinfo{author}{Porf\'\i{}rio, P.J.},
  \bibinfo{year}{2024}.
\newblock \bibinfo{title}{{An exact stationary axisymmetric vacuum solution
  within a metric-affine bumblebee gravity}}.
\newblock \bibinfo{journal}{JCAP} \bibinfo{volume}{07}, \bibinfo{pages}{004}.
\newblock \DOIprefix\doi{10.1088/1475-7516/2024/07/004},
  \href{http://arxiv.org/abs/2402.13014}{{\tt arXiv:2402.13014}}.
\bibitem[{Askins et~al.(2020)}]{Theia:2019non}
\bibinfo{author}{Askins, M.}, et~al. (\bibinfo{collaboration}{Theia}),
  \bibinfo{year}{2020}.
\newblock \bibinfo{title}{{THEIA: an advanced optical neutrino detector}}.
\newblock \bibinfo{journal}{Eur. Phys. J. C} \bibinfo{volume}{80},
  \bibinfo{pages}{416}.
\newblock \DOIprefix\doi{10.1140/epjc/s10052-020-7977-8},
  \href{http://arxiv.org/abs/1911.03501}{{\tt arXiv:1911.03501}}.
\bibitem[{Auclair et~al.(2023)Auclair, Blasi, Brdar and
  Schmitz}]{Auclair:2022ylu}
\bibinfo{author}{Auclair, P.}, \bibinfo{author}{Blasi, S.},
  \bibinfo{author}{Brdar, V.}, \bibinfo{author}{Schmitz, K.},
  \bibinfo{year}{2023}.
\newblock \bibinfo{title}{{Gravitational waves from current-carrying cosmic
  strings}}.
\newblock \bibinfo{journal}{JCAP} \bibinfo{volume}{04}, \bibinfo{pages}{009}.
\newblock \DOIprefix\doi{10.1088/1475-7516/2023/04/009},
  \href{http://arxiv.org/abs/2207.03510}{{\tt arXiv:2207.03510}}.
\bibitem[{Badurina et~al.(2020)}]{Badurina:2019hst}
\bibinfo{author}{Badurina, L.}, et~al., \bibinfo{year}{2020}.
\newblock \bibinfo{title}{{AION: An Atom Interferometer Observatory and
  Network}}.
\newblock \bibinfo{journal}{JCAP} \bibinfo{volume}{05}, \bibinfo{pages}{011}.
\newblock \DOIprefix\doi{10.1088/1475-7516/2020/05/011},
  \href{http://arxiv.org/abs/1911.11755}{{\tt arXiv:1911.11755}}.
\bibitem[{Battye et~al.(1998)Battye, Caldwell and Shellard}]{Battye1998}
\bibinfo{author}{Battye, R.A.}, \bibinfo{author}{Caldwell, R.R.},
  \bibinfo{author}{Shellard, E.P.S.}, \bibinfo{year}{1998}.
\newblock \bibinfo{title}{Gravitational waves from cosmic strings}, in:
  \bibinfo{booktitle}{Topological Defects in Cosmology}.
\bibitem[{Bernal et~al.(2020)Bernal, Ghoshal, Hajkarim and
  Lambiase}]{Bernal:2020ywq}
\bibinfo{author}{Bernal, N.}, \bibinfo{author}{Ghoshal, A.},
  \bibinfo{author}{Hajkarim, F.}, \bibinfo{author}{Lambiase, G.},
  \bibinfo{year}{2020}.
\newblock \bibinfo{title}{{Primordial Gravitational Wave Signals in Modified
  Cosmologies}}.
\newblock \bibinfo{journal}{JCAP} \bibinfo{volume}{11}, \bibinfo{pages}{051}.
\newblock \DOIprefix\doi{10.1088/1475-7516/2020/11/051},
  \href{http://arxiv.org/abs/2008.04959}{{\tt arXiv:2008.04959}}.
\bibitem[{Bernal and Hajkarim(2019)}]{Bernal:2019lpc}
\bibinfo{author}{Bernal, N.}, \bibinfo{author}{Hajkarim, F.},
  \bibinfo{year}{2019}.
\newblock \bibinfo{title}{{Primordial Gravitational Waves in Nonstandard
  Cosmologies}}.
\newblock \bibinfo{journal}{Phys. Rev. D} \bibinfo{volume}{100},
  \bibinfo{pages}{063502}.
\newblock \DOIprefix\doi{10.1103/PhysRevD.100.063502},
  \href{http://arxiv.org/abs/1905.10410}{{\tt arXiv:1905.10410}}.
\bibitem[{Bertolami and Paramos(2005)}]{Bertolami:2005bh}
\bibinfo{author}{Bertolami, O.}, \bibinfo{author}{Paramos, J.},
  \bibinfo{year}{2005}.
\newblock \bibinfo{title}{{The Flight of the bumblebee: Vacuum solutions of a
  gravity model with vector-induced spontaneous Lorentz symmetry breaking}}.
\newblock \bibinfo{journal}{Phys. Rev. D} \bibinfo{volume}{72},
  \bibinfo{pages}{044001}.
\newblock \DOIprefix\doi{10.1103/PhysRevD.72.044001},
  \href{http://arxiv.org/abs/hep-th/0504215}{{\tt arXiv:hep-th/0504215}}.
\bibitem[{Bjorken(1963)}]{Bjorken:1963vg}
\bibinfo{author}{Bjorken, J.D.}, \bibinfo{year}{1963}.
\newblock \bibinfo{title}{{A Dynamical origin for the electromagnetic field}}.
\newblock \bibinfo{journal}{Annals Phys.} \bibinfo{volume}{24},
  \bibinfo{pages}{174--187}.
\newblock \DOIprefix\doi{10.1016/0003-4916(63)90069-1}.
\bibitem[{Bluhm and Kostelecky(2005)}]{Bluhm:2004ep}
\bibinfo{author}{Bluhm, R.}, \bibinfo{author}{Kostelecky, V.A.},
  \bibinfo{year}{2005}.
\newblock \bibinfo{title}{{Spontaneous Lorentz violation, Nambu-Goldstone
  modes, and gravity}}.
\newblock \bibinfo{journal}{Phys. Rev. D} \bibinfo{volume}{71},
  \bibinfo{pages}{065008}.
\newblock \DOIprefix\doi{10.1103/PhysRevD.71.065008},
  \href{http://arxiv.org/abs/hep-th/0412320}{{\tt arXiv:hep-th/0412320}}.
\bibitem[{Boyle and Buonanno(2008)}]{Boyle:2007zx}
\bibinfo{author}{Boyle, L.A.}, \bibinfo{author}{Buonanno, A.},
  \bibinfo{year}{2008}.
\newblock \bibinfo{title}{{Relating gravitational wave constraints from
  primordial nucleosynthesis, pulsar timing, laser interferometers, and the
  CMB: Implications for the early Universe}}.
\newblock \bibinfo{journal}{Phys. Rev. D} \bibinfo{volume}{78},
  \bibinfo{pages}{043531}.
\newblock \DOIprefix\doi{10.1103/PhysRevD.78.043531},
  \href{http://arxiv.org/abs/0708.2279}{{\tt arXiv:0708.2279}}.
\bibitem[{Braglia et~al.(2024)}]{LISACosmologyWorkingGroup:2024hsc}
\bibinfo{author}{Braglia, M.}, et~al. (\bibinfo{collaboration}{LISA Cosmology
  Working Group}), \bibinfo{year}{2024}.
\newblock \bibinfo{title}{{Gravitational waves from inflation in LISA:
  reconstruction pipeline and physics interpretation}}.
\newblock \bibinfo{journal}{JCAP} \bibinfo{volume}{11}, \bibinfo{pages}{032}.
\newblock \DOIprefix\doi{10.1088/1475-7516/2024/11/032},
  \href{http://arxiv.org/abs/2407.04356}{{\tt arXiv:2407.04356}}.
\bibitem[{Brandenburg et~al.(2021)Brandenburg, Clarke, He and
  Kahniashvili}]{Brandenburg:2021tmp}
\bibinfo{author}{Brandenburg, A.}, \bibinfo{author}{Clarke, E.},
  \bibinfo{author}{He, Y.}, \bibinfo{author}{Kahniashvili, T.},
  \bibinfo{year}{2021}.
\newblock \bibinfo{title}{{Can we observe the QCD phase transition-generated
  gravitational waves through pulsar timing arrays?}}
\newblock \bibinfo{journal}{Phys. Rev. D} \bibinfo{volume}{104},
  \bibinfo{pages}{043513}.
\newblock \DOIprefix\doi{10.1103/PhysRevD.104.043513},
  \href{http://arxiv.org/abs/2102.12428}{{\tt arXiv:2102.12428}}.
\bibitem[{Capelo and P\'aramos(2015)}]{Capelo:2015ipa}
\bibinfo{author}{Capelo, D.}, \bibinfo{author}{P\'aramos, J.},
  \bibinfo{year}{2015}.
\newblock \bibinfo{title}{{Cosmological implications of Bumblebee vector
  models}}.
\newblock \bibinfo{journal}{Phys. Rev. D} \bibinfo{volume}{91},
  \bibinfo{pages}{104007}.
\newblock \DOIprefix\doi{10.1103/PhysRevD.91.104007},
  \href{http://arxiv.org/abs/1501.07685}{{\tt arXiv:1501.07685}}.
\bibitem[{Capozziello et~al.(2019)Capozziello, Khodadi and
  Lambiase}]{Capozziello:2018qjs}
\bibinfo{author}{Capozziello, S.}, \bibinfo{author}{Khodadi, M.},
  \bibinfo{author}{Lambiase, G.}, \bibinfo{year}{2019}.
\newblock \bibinfo{title}{{The quark chemical potential of QCD phase transition
  and the stochastic background of gravitational waves}}.
\newblock \bibinfo{journal}{Phys. Lett. B} \bibinfo{volume}{789},
  \bibinfo{pages}{626--633}.
\newblock \DOIprefix\doi{10.1016/j.physletb.2019.01.004},
  \href{http://arxiv.org/abs/1808.06188}{{\tt arXiv:1808.06188}}.
\bibitem[{Capozziello et~al.(2023a)Capozziello, Zare, Mota and
  Hassanabadi}]{Capozziello:2023rfv}
\bibinfo{author}{Capozziello, S.}, \bibinfo{author}{Zare, S.},
  \bibinfo{author}{Mota, D.F.}, \bibinfo{author}{Hassanabadi, H.},
  \bibinfo{year}{2023}a.
\newblock \bibinfo{title}{{Dark matter spike around Bumblebee black holes}}
  \DOIprefix\doi{10.1088/1475-7516/2023/05/027},
  \href{http://arxiv.org/abs/2303.13554}{{\tt arXiv:2303.13554}}.
\bibitem[{Capozziello et~al.(2023b)Capozziello, Zare, Nieto and
  Hassanabadi}]{Capozziello:2023tbo}
\bibinfo{author}{Capozziello, S.}, \bibinfo{author}{Zare, S.},
  \bibinfo{author}{Nieto, L.M.}, \bibinfo{author}{Hassanabadi, H.},
  \bibinfo{year}{2023}b.
\newblock \bibinfo{title}{{Modified Kerr black holes surrounded by dark matter
  spike}} \href{http://arxiv.org/abs/2311.12896}{{\tt arXiv:2311.12896}}.
\bibitem[{Caprini et~al.(2010)Caprini, Durrer and Siemens}]{Caprini:2010xv}
\bibinfo{author}{Caprini, C.}, \bibinfo{author}{Durrer, R.},
  \bibinfo{author}{Siemens, X.}, \bibinfo{year}{2010}.
\newblock \bibinfo{title}{{Detection of gravitational waves from the QCD phase
  transition with pulsar timing arrays}}.
\newblock \bibinfo{journal}{Phys. Rev. D} \bibinfo{volume}{82},
  \bibinfo{pages}{063511}.
\newblock \DOIprefix\doi{10.1103/PhysRevD.82.063511},
  \href{http://arxiv.org/abs/1007.1218}{{\tt arXiv:1007.1218}}.
\bibitem[{Carroll et~al.(2001)Carroll, Harvey, Kostelecky, Lane and
  Okamoto}]{Carroll:2001ws}
\bibinfo{author}{Carroll, S.M.}, \bibinfo{author}{Harvey, J.A.},
  \bibinfo{author}{Kostelecky, V.A.}, \bibinfo{author}{Lane, C.D.},
  \bibinfo{author}{Okamoto, T.}, \bibinfo{year}{2001}.
\newblock \bibinfo{title}{{Noncommutative field theory and Lorentz violation}}.
\newblock \bibinfo{journal}{Phys. Rev. Lett.} \bibinfo{volume}{87},
  \bibinfo{pages}{141601}.
\newblock \DOIprefix\doi{10.1103/PhysRevLett.87.141601},
  \href{http://arxiv.org/abs/hep-th/0105082}{{\tt arXiv:hep-th/0105082}}.
\bibitem[{Casana et~al.(2018)Casana, Cavalcante, Poulis and
  Santos}]{Casana:2017jkc}
\bibinfo{author}{Casana, R.}, \bibinfo{author}{Cavalcante, A.},
  \bibinfo{author}{Poulis, F.P.}, \bibinfo{author}{Santos, E.B.},
  \bibinfo{year}{2018}.
\newblock \bibinfo{title}{{Exact Schwarzschild-like solution in a bumblebee
  gravity model}}.
\newblock \bibinfo{journal}{Phys. Rev. D} \bibinfo{volume}{97},
  \bibinfo{pages}{104001}.
\newblock \DOIprefix\doi{10.1103/PhysRevD.97.104001},
  \href{http://arxiv.org/abs/1711.02273}{{\tt arXiv:1711.02273}}.
\bibitem[{Chamseddine and Mukhanov(2013)}]{Chamseddine:2013kea}
\bibinfo{author}{Chamseddine, A.H.}, \bibinfo{author}{Mukhanov, V.},
  \bibinfo{year}{2013}.
\newblock \bibinfo{title}{{Mimetic Dark Matter}}.
\newblock \bibinfo{journal}{JHEP} \bibinfo{volume}{11}, \bibinfo{pages}{135}.
\newblock \DOIprefix\doi{10.1007/JHEP11(2013)135},
  \href{http://arxiv.org/abs/1308.5410}{{\tt arXiv:1308.5410}}.
\bibitem[{Chang and Cui(2022)}]{Chang:2021afa}
\bibinfo{author}{Chang, C.F.}, \bibinfo{author}{Cui, Y.}, \bibinfo{year}{2022}.
\newblock \bibinfo{title}{{Gravitational waves from global cosmic strings and
  cosmic archaeology}}.
\newblock \bibinfo{journal}{JHEP} \bibinfo{volume}{03}, \bibinfo{pages}{114}.
\newblock \DOIprefix\doi{10.1007/JHEP03(2022)114},
  \href{http://arxiv.org/abs/2106.09746}{{\tt arXiv:2106.09746}}.
\bibitem[{Chen et~al.(2018)Chen, Huang and Yan}]{Chen:2017cyc}
\bibinfo{author}{Chen, Y.}, \bibinfo{author}{Huang, M.}, \bibinfo{author}{Yan,
  Q.S.}, \bibinfo{year}{2018}.
\newblock \bibinfo{title}{{Gravitation waves from QCD and electroweak phase
  transitions}}.
\newblock \bibinfo{journal}{JHEP} \bibinfo{volume}{05}, \bibinfo{pages}{178}.
\newblock \DOIprefix\doi{10.1007/JHEP05(2018)178},
  \href{http://arxiv.org/abs/1712.03470}{{\tt arXiv:1712.03470}}.
\bibitem[{Clarke et~al.(2020)Clarke, Copeland and Moss}]{Clarke:2020bil}
\bibinfo{author}{Clarke, T.J.}, \bibinfo{author}{Copeland, E.J.},
  \bibinfo{author}{Moss, A.}, \bibinfo{year}{2020}.
\newblock \bibinfo{title}{{Constraints on primordial gravitational waves from
  the Cosmic Microwave Background}}.
\newblock \bibinfo{journal}{JCAP} \bibinfo{volume}{10}, \bibinfo{pages}{002}.
\newblock \DOIprefix\doi{10.1088/1475-7516/2020/10/002},
  \href{http://arxiv.org/abs/2004.11396}{{\tt arXiv:2004.11396}}.
\bibitem[{Colladay and Kostelecky(1997)}]{Colladay:1996iz}
\bibinfo{author}{Colladay, D.}, \bibinfo{author}{Kostelecky, V.A.},
  \bibinfo{year}{1997}.
\newblock \bibinfo{title}{{CPT violation and the standard model}}.
\newblock \bibinfo{journal}{Phys. Rev. D} \bibinfo{volume}{55},
  \bibinfo{pages}{6760--6774}.
\newblock \DOIprefix\doi{10.1103/PhysRevD.55.6760},
  \href{http://arxiv.org/abs/hep-ph/9703464}{{\tt arXiv:hep-ph/9703464}}.
\bibitem[{Colladay and Kostelecky(1998)}]{Colladay:1998fq}
\bibinfo{author}{Colladay, D.}, \bibinfo{author}{Kostelecky, V.A.},
  \bibinfo{year}{1998}.
\newblock \bibinfo{title}{{Lorentz violating extension of the standard model}}.
\newblock \bibinfo{journal}{Phys. Rev. D} \bibinfo{volume}{58},
  \bibinfo{pages}{116002}.
\newblock \DOIprefix\doi{10.1103/PhysRevD.58.116002},
  \href{http://arxiv.org/abs/hep-ph/9809521}{{\tt arXiv:hep-ph/9809521}}.
\bibitem[{Crowder and Cornish(2005)}]{Crowder:2005nr}
\bibinfo{author}{Crowder, J.}, \bibinfo{author}{Cornish, N.J.},
  \bibinfo{year}{2005}.
\newblock \bibinfo{title}{{Beyond LISA: Exploring future gravitational wave
  missions}}.
\newblock \bibinfo{journal}{Phys. Rev. D} \bibinfo{volume}{72},
  \bibinfo{pages}{083005}.
\newblock \DOIprefix\doi{10.1103/PhysRevD.72.083005},
  \href{http://arxiv.org/abs/gr-qc/0506015}{{\tt arXiv:gr-qc/0506015}}.
\bibitem[{Cutting et~al.(2018)Cutting, Hindmarsh and Weir}]{Cutting:2018tjt}
\bibinfo{author}{Cutting, D.}, \bibinfo{author}{Hindmarsh, M.},
  \bibinfo{author}{Weir, D.J.}, \bibinfo{year}{2018}.
\newblock \bibinfo{title}{{Gravitational waves from vacuum first-order phase
  transitions: from the envelope to the lattice}}.
\newblock \bibinfo{journal}{Phys. Rev. D} \bibinfo{volume}{97},
  \bibinfo{pages}{123513}.
\newblock \DOIprefix\doi{10.1103/PhysRevD.97.123513},
  \href{http://arxiv.org/abs/1802.05712}{{\tt arXiv:1802.05712}}.
\bibitem[{Dai and Stojkovic(2019)}]{Dai:2019ksi}
\bibinfo{author}{Dai, D.C.}, \bibinfo{author}{Stojkovic, D.},
  \bibinfo{year}{2019}.
\newblock \bibinfo{title}{{Primordial scalar gravitational waves produced at
  the QCD phase transition due to the trace anomaly}}.
\newblock \bibinfo{journal}{Class. Quant. Grav.} \bibinfo{volume}{36},
  \bibinfo{pages}{145004}.
\newblock \DOIprefix\doi{10.1088/1361-6382/ab2288},
  \href{http://arxiv.org/abs/1905.05850}{{\tt arXiv:1905.05850}}.
\bibitem[{Davoudiasl(2019)}]{Davoudiasl:2019ugw}
\bibinfo{author}{Davoudiasl, H.}, \bibinfo{year}{2019}.
\newblock \bibinfo{title}{{LIGO/Virgo Black Holes from a First Order Quark
  Confinement Phase Transition}}.
\newblock \bibinfo{journal}{Phys. Rev. Lett.} \bibinfo{volume}{123},
  \bibinfo{pages}{101102}.
\newblock \DOIprefix\doi{10.1103/PhysRevLett.123.101102},
  \href{http://arxiv.org/abs/1902.07805}{{\tt arXiv:1902.07805}}.
\bibitem[{Delhom et~al.(2021)Delhom, Nascimento, Olmo, Petrov and
  Porf\'\i{}rio}]{Delhom:2019wcm}
\bibinfo{author}{Delhom, A.}, \bibinfo{author}{Nascimento, J.R.},
  \bibinfo{author}{Olmo, G.J.}, \bibinfo{author}{Petrov, A.Y.},
  \bibinfo{author}{Porf\'\i{}rio, P.J.}, \bibinfo{year}{2021}.
\newblock \bibinfo{title}{{Metric-affine bumblebee gravity: classical
  aspects}}.
\newblock \bibinfo{journal}{Eur. Phys. J. C} \bibinfo{volume}{81},
  \bibinfo{pages}{287}.
\newblock \DOIprefix\doi{10.1140/epjc/s10052-021-09083-y},
  \href{http://arxiv.org/abs/1911.11605}{{\tt arXiv:1911.11605}}.
\bibitem[{Dent et~al.(2013)Dent, Krauss, Sabharwal and
  Vachaspati}]{Dent:2013asa}
\bibinfo{author}{Dent, J.B.}, \bibinfo{author}{Krauss, L.M.},
  \bibinfo{author}{Sabharwal, S.}, \bibinfo{author}{Vachaspati, T.},
  \bibinfo{year}{2013}.
\newblock \bibinfo{title}{{Damping of Primordial Gravitational Waves from
  Generalized Sources}}.
\newblock \bibinfo{journal}{Phys. Rev. D} \bibinfo{volume}{88},
  \bibinfo{pages}{084008}.
\newblock \DOIprefix\doi{10.1103/PhysRevD.88.084008},
  \href{http://arxiv.org/abs/1307.7571}{{\tt arXiv:1307.7571}}.
\bibitem[{Ding et~al.(2020)Ding, Liu, Casana and Cavalcante}]{Ding:2019mal}
\bibinfo{author}{Ding, C.}, \bibinfo{author}{Liu, C.}, \bibinfo{author}{Casana,
  R.}, \bibinfo{author}{Cavalcante, A.}, \bibinfo{year}{2020}.
\newblock \bibinfo{title}{{Exact Kerr-like solution and its shadow in a gravity
  model with spontaneous Lorentz symmetry breaking}}.
\newblock \bibinfo{journal}{Eur. Phys. J. C} \bibinfo{volume}{80},
  \bibinfo{pages}{178}.
\newblock \DOIprefix\doi{10.1140/epjc/s10052-020-7743-y},
  \href{http://arxiv.org/abs/1910.02674}{{\tt arXiv:1910.02674}}.
\bibitem[{Ding et~al.(2024)Ding, Liu, Xiao and Chen}]{Ding:2024qrf}
\bibinfo{author}{Ding, C.}, \bibinfo{author}{Liu, C.}, \bibinfo{author}{Xiao,
  Y.}, \bibinfo{author}{Chen, J.}, \bibinfo{year}{2024}.
\newblock \bibinfo{title}{{Phantom hairy black holes and wormholes in
  Einstein-bumblebee gravity}} \href{http://arxiv.org/abs/2407.16916}{{\tt
  arXiv:2407.16916}}.
\bibitem[{Ding et~al.(2023)Ding, Shi, Chen, Zhou, Liu and Xiao}]{Ding:2023niy}
\bibinfo{author}{Ding, C.}, \bibinfo{author}{Shi, Y.}, \bibinfo{author}{Chen,
  J.}, \bibinfo{author}{Zhou, Y.}, \bibinfo{author}{Liu, C.},
  \bibinfo{author}{Xiao, Y.}, \bibinfo{year}{2023}.
\newblock \bibinfo{title}{{Rotating BTZ-like black hole and central charges in
  Einstein-bumblebee gravity}}.
\newblock \bibinfo{journal}{Eur. Phys. J. C} \bibinfo{volume}{83},
  \bibinfo{pages}{573}.
\newblock \DOIprefix\doi{10.1140/epjc/s10052-023-11761-y},
  \href{http://arxiv.org/abs/2302.01580}{{\tt arXiv:2302.01580}}.
\bibitem[{DIRAC(1951)}]{Dirac1951}
\bibinfo{author}{DIRAC, P.A.M.}, \bibinfo{year}{1951}.
\newblock \bibinfo{title}{Is there an {\ae}ther?}
\newblock \bibinfo{journal}{Nature} \bibinfo{volume}{168},
  \bibinfo{pages}{906--907}.
\newblock \URLprefix \url{https://doi.org/10.1038/168906a0},
  \DOIprefix\doi{10.1038/168906a0}.
\bibitem[{Dodelson(2003)}]{Dodelson:2003ft}
\bibinfo{author}{Dodelson, S.}, \bibinfo{year}{2003}.
\newblock \bibinfo{title}{{Modern Cosmology}}.
\newblock \bibinfo{publisher}{Academic Press}, \bibinfo{address}{Amsterdam}.
\bibitem[{El-Neaj et~al.(2020)}]{AEDGE:2019nxb}
\bibinfo{author}{El-Neaj, Y.A.}, et~al. (\bibinfo{collaboration}{AEDGE}),
  \bibinfo{year}{2020}.
\newblock \bibinfo{title}{{AEDGE: Atomic Experiment for Dark Matter and Gravity
  Exploration in Space}}.
\newblock \bibinfo{journal}{EPJ Quant. Technol.} \bibinfo{volume}{7},
  \bibinfo{pages}{6}.
\newblock \DOIprefix\doi{10.1140/epjqt/s40507-020-0080-0},
  \href{http://arxiv.org/abs/1908.00802}{{\tt arXiv:1908.00802}}.
\bibitem[{Eling et~al.(2004)Eling, Jacobson and Mattingly}]{Eling:2004dk}
\bibinfo{author}{Eling, C.}, \bibinfo{author}{Jacobson, T.},
  \bibinfo{author}{Mattingly, D.}, \bibinfo{year}{2004}.
\newblock \bibinfo{title}{{Einstein-Aether theory}}, in:
  \bibinfo{booktitle}{{Deserfest: A Celebration of the Life and Works of
  Stanley Deser}}, pp. \bibinfo{pages}{163--179}.
\newblock \href{http://arxiv.org/abs/gr-qc/0410001}{{\tt arXiv:gr-qc/0410001}}.
\bibitem[{Ellis and Stoeger(1987)}]{Ellis:1987zz}
\bibinfo{author}{Ellis, G.F.R.}, \bibinfo{author}{Stoeger, W.},
  \bibinfo{year}{1987}.
\newblock \bibinfo{title}{{The 'fitting problem' in cosmology}}.
\newblock \bibinfo{journal}{Class. Quant. Grav.} \bibinfo{volume}{4},
  \bibinfo{pages}{1697--1729}.
\newblock \DOIprefix\doi{10.1088/0264-9381/4/6/025}.
\bibitem[{Evans et~al.(2021)}]{Evans:2021gyd}
\bibinfo{author}{Evans, M.}, et~al., \bibinfo{year}{2021}.
\newblock \bibinfo{title}{{A Horizon Study for Cosmic Explorer: Science,
  Observatories, and Community}} \href{http://arxiv.org/abs/2109.09882}{{\tt
  arXiv:2109.09882}}.
\bibitem[{Feng et~al.(2023)Feng, Feng, Zhou and Jiang}]{Feng:2022fwf}
\bibinfo{author}{Feng, Q.M.}, \bibinfo{author}{Feng, Z.W.},
  \bibinfo{author}{Zhou, X.}, \bibinfo{author}{Jiang, Q.Q.},
  \bibinfo{year}{2023}.
\newblock \bibinfo{title}{{Barrow entropy and stochastic gravitational wave
  background generated from cosmological QCD phase transition}}.
\newblock \bibinfo{journal}{Phys. Lett. B} \bibinfo{volume}{838},
  \bibinfo{pages}{137739}.
\newblock \DOIprefix\doi{10.1016/j.physletb.2023.137739},
  \href{http://arxiv.org/abs/2210.10658}{{\tt arXiv:2210.10658}}.
\bibitem[{Filho et~al.(2023)Filho, Nascimento, Petrov and
  Porf\'\i{}rio}]{Filho:2022yrk}
\bibinfo{author}{Filho, A.A.A.}, \bibinfo{author}{Nascimento, J.R.},
  \bibinfo{author}{Petrov, A.Y.}, \bibinfo{author}{Porf\'\i{}rio, P.J.},
  \bibinfo{year}{2023}.
\newblock \bibinfo{title}{{Vacuum solution within a metric-affine bumblebee
  gravity}}.
\newblock \bibinfo{journal}{Phys. Rev. D} \bibinfo{volume}{108},
  \bibinfo{pages}{085010}.
\newblock \DOIprefix\doi{10.1103/PhysRevD.108.085010},
  \href{http://arxiv.org/abs/2211.11821}{{\tt arXiv:2211.11821}}.
\bibitem[{Ghosh et~al.(2023)Ghosh, Nair, Pathak, Sarkar and
  Sengupta}]{Ghosh:2023xes}
\bibinfo{author}{Ghosh, R.}, \bibinfo{author}{Nair, S.},
  \bibinfo{author}{Pathak, L.}, \bibinfo{author}{Sarkar, S.},
  \bibinfo{author}{Sengupta, A.S.}, \bibinfo{year}{2023}.
\newblock \bibinfo{title}{{Does the speed of gravitational waves depend on the
  source velocity?}}
\newblock \bibinfo{journal}{Phys. Rev. D} \bibinfo{volume}{108},
  \bibinfo{pages}{124017}.
\newblock \DOIprefix\doi{10.1103/PhysRevD.108.124017},
  \href{http://arxiv.org/abs/2304.14820}{{\tt arXiv:2304.14820}}.
\bibitem[{Gon\c{c}alves et~al.(2022)Gon\c{c}alves, Kaladharan and
  Wu}]{Goncalves:2021egx}
\bibinfo{author}{Gon\c{c}alves, D.}, \bibinfo{author}{Kaladharan, A.},
  \bibinfo{author}{Wu, Y.}, \bibinfo{year}{2022}.
\newblock \bibinfo{title}{{Electroweak phase transition in the 2HDM: Collider
  and gravitational wave complementarity}}.
\newblock \bibinfo{journal}{Phys. Rev. D} \bibinfo{volume}{105},
  \bibinfo{pages}{095041}.
\newblock \DOIprefix\doi{10.1103/PhysRevD.105.095041},
  \href{http://arxiv.org/abs/2108.05356}{{\tt arXiv:2108.05356}}.
\bibitem[{Green and Wald(2014)}]{Green:2014aga}
\bibinfo{author}{Green, S.R.}, \bibinfo{author}{Wald, R.M.},
  \bibinfo{year}{2014}.
\newblock \bibinfo{title}{{How well is our universe described by an FLRW
  model?}}
\newblock \bibinfo{journal}{Class. Quant. Grav.} \bibinfo{volume}{31},
  \bibinfo{pages}{234003}.
\newblock \DOIprefix\doi{10.1088/0264-9381/31/23/234003},
  \href{http://arxiv.org/abs/1407.8084}{{\tt arXiv:1407.8084}}.
\bibitem[{G\"ull\"u and \"Ovg\"un(2022)}]{Gullu:2020qzu}
\bibinfo{author}{G\"ull\"u, I.}, \bibinfo{author}{\"Ovg\"un, A.},
  \bibinfo{year}{2022}.
\newblock \bibinfo{title}{{Schwarzschild-like black hole with a topological
  defect in bumblebee gravity}}.
\newblock \bibinfo{journal}{Annals Phys.} \bibinfo{volume}{436},
  \bibinfo{pages}{168721}.
\newblock \DOIprefix\doi{10.1016/j.aop.2021.168721},
  \href{http://arxiv.org/abs/2012.02611}{{\tt arXiv:2012.02611}}.
\bibitem[{Guth(1981)}]{Guth:1980zm}
\bibinfo{author}{Guth, A.H.}, \bibinfo{year}{1981}.
\newblock \bibinfo{title}{{The Inflationary Universe: A Possible Solution to
  the Horizon and Flatness Problems}}.
\newblock \bibinfo{journal}{Phys. Rev. D} \bibinfo{volume}{23},
  \bibinfo{pages}{347--356}.
\newblock \DOIprefix\doi{10.1103/PhysRevD.23.347}.
\bibitem[{Guzzetti et~al.(2016)Guzzetti, Bartolo, Liguori and
  Matarrese}]{Guzzetti:2016mkm}
\bibinfo{author}{Guzzetti, M.C.}, \bibinfo{author}{Bartolo, N.},
  \bibinfo{author}{Liguori, M.}, \bibinfo{author}{Matarrese, S.},
  \bibinfo{year}{2016}.
\newblock \bibinfo{title}{{Gravitational waves from inflation}}.
\newblock \bibinfo{journal}{Riv. Nuovo Cim.} \bibinfo{volume}{39},
  \bibinfo{pages}{399--495}.
\newblock \DOIprefix\doi{10.1393/ncr/i2016-10127-1},
  \href{http://arxiv.org/abs/1605.01615}{{\tt arXiv:1605.01615}}.
\bibitem[{Hajkarim et~al.(2019)Hajkarim, Schaffner-Bielich, Wystub and
  Wygas}]{Hajkarim:2019csy}
\bibinfo{author}{Hajkarim, F.}, \bibinfo{author}{Schaffner-Bielich, J.},
  \bibinfo{author}{Wystub, S.}, \bibinfo{author}{Wygas, M.M.},
  \bibinfo{year}{2019}.
\newblock \bibinfo{title}{{Effects of the QCD Equation of State and Lepton
  Asymmetry on Primordial Gravitational Waves}}.
\newblock \bibinfo{journal}{Phys. Rev. D} \bibinfo{volume}{99},
  \bibinfo{pages}{103527}.
\newblock \DOIprefix\doi{10.1103/PhysRevD.99.103527},
  \href{http://arxiv.org/abs/1904.01046}{{\tt arXiv:1904.01046}}.
\bibitem[{Horava(2009)}]{Horava:2009uw}
\bibinfo{author}{Horava, P.}, \bibinfo{year}{2009}.
\newblock \bibinfo{title}{{Quantum Gravity at a Lifshitz Point}}.
\newblock \bibinfo{journal}{Phys. Rev. D} \bibinfo{volume}{79},
  \bibinfo{pages}{084008}.
\newblock \DOIprefix\doi{10.1103/PhysRevD.79.084008},
  \href{http://arxiv.org/abs/0901.3775}{{\tt arXiv:0901.3775}}.
\bibitem[{Janssen et~al.(2015)}]{Janssen:2014dka}
\bibinfo{author}{Janssen, G.}, et~al., \bibinfo{year}{2015}.
\newblock \bibinfo{title}{{Gravitational wave astronomy with the SKA}}.
\newblock \bibinfo{journal}{PoS} \bibinfo{volume}{AASKA14},
  \bibinfo{pages}{037}.
\newblock \DOIprefix\doi{10.22323/1.215.0037},
  \href{http://arxiv.org/abs/1501.00127}{{\tt arXiv:1501.00127}}.
\bibitem[{Jesus and Santos(2019)}]{Jesus:2019nwi}
\bibinfo{author}{Jesus, W.D.R.}, \bibinfo{author}{Santos, A.F.},
  \bibinfo{year}{2019}.
\newblock \bibinfo{title}{{Ricci dark energy in bumblebee gravity model}}.
\newblock \bibinfo{journal}{Mod. Phys. Lett. A} \bibinfo{volume}{34},
  \bibinfo{pages}{1950171}.
\newblock \DOIprefix\doi{10.1142/S0217732319501712},
  \href{http://arxiv.org/abs/1903.09316}{{\tt arXiv:1903.09316}}.
\bibitem[{Jesus and Santos(2020)}]{Jesus:2020lsv}
\bibinfo{author}{Jesus, W.D.R.}, \bibinfo{author}{Santos, A.F.},
  \bibinfo{year}{2020}.
\newblock \bibinfo{title}{{G\"odel-type universes in bumblebee gravity}}.
\newblock \bibinfo{journal}{Int. J. Mod. Phys. A} \bibinfo{volume}{35},
  \bibinfo{pages}{2050050}.
\newblock \DOIprefix\doi{10.1142/S0217751X20500505},
  \href{http://arxiv.org/abs/2003.13364}{{\tt arXiv:2003.13364}}.
\bibitem[{Jha et~al.(2022)Jha, Aziz and Rahaman}]{Jha:2021eww}
\bibinfo{author}{Jha, S.K.}, \bibinfo{author}{Aziz, S.},
  \bibinfo{author}{Rahaman, A.}, \bibinfo{year}{2022}.
\newblock \bibinfo{title}{{Study of Einstein-bumblebee gravity with
  Kerr-Sen-like solution in the presence of a dispersive medium}}.
\newblock \bibinfo{journal}{Eur. Phys. J. C} \bibinfo{volume}{82},
  \bibinfo{pages}{106}.
\newblock \DOIprefix\doi{10.1140/epjc/s10052-022-10042-4},
  \href{http://arxiv.org/abs/2103.17021}{{\tt arXiv:2103.17021}}.
\bibitem[{Jha and Rahaman(2021)}]{Jha:2020pvk}
\bibinfo{author}{Jha, S.K.}, \bibinfo{author}{Rahaman, A.},
  \bibinfo{year}{2021}.
\newblock \bibinfo{title}{{Bumblebee gravity with a Kerr-Sen-like solution and
  its Shadow}}.
\newblock \bibinfo{journal}{Eur. Phys. J. C} \bibinfo{volume}{81},
  \bibinfo{pages}{345}.
\newblock \DOIprefix\doi{10.1140/epjc/s10052-021-09132-6},
  \href{http://arxiv.org/abs/2011.14916}{{\tt arXiv:2011.14916}}.
\bibitem[{Ji et~al.(2024)Ji, Li, Yang, Xu, Hu and Shao}]{Ji:2024aeg}
\bibinfo{author}{Ji, P.}, \bibinfo{author}{Li, Z.}, \bibinfo{author}{Yang, L.},
  \bibinfo{author}{Xu, R.}, \bibinfo{author}{Hu, Z.}, \bibinfo{author}{Shao,
  L.}, \bibinfo{year}{2024}.
\newblock \bibinfo{title}{{Neutron stars in the bumblebee theory of gravity}}.
\newblock \bibinfo{journal}{Phys. Rev. D} \bibinfo{volume}{110},
  \bibinfo{pages}{104057}.
\newblock \DOIprefix\doi{10.1103/PhysRevD.110.104057},
  \href{http://arxiv.org/abs/2409.04805}{{\tt arXiv:2409.04805}}.
\bibitem[{Jizba et~al.(2024)Jizba, Lambiase, Luciano and
  Mastrototaro}]{Jizba:2024klq}
\bibinfo{author}{Jizba, P.}, \bibinfo{author}{Lambiase, G.},
  \bibinfo{author}{Luciano, G.G.}, \bibinfo{author}{Mastrototaro, L.},
  \bibinfo{year}{2024}.
\newblock \bibinfo{title}{{Imprints of Barrow\textendash{}Tsallis cosmology in
  primordial gravitational waves}}.
\newblock \bibinfo{journal}{Eur. Phys. J. C} \bibinfo{volume}{84},
  \bibinfo{pages}{1076}.
\newblock \DOIprefix\doi{10.1140/epjc/s10052-024-13455-5},
  \href{http://arxiv.org/abs/2403.09797}{{\tt arXiv:2403.09797}}.
\bibitem[{Kamionkowski et~al.(1997)Kamionkowski, Kosowsky and
  Stebbins}]{Kamionkowski:1996ks}
\bibinfo{author}{Kamionkowski, M.}, \bibinfo{author}{Kosowsky, A.},
  \bibinfo{author}{Stebbins, A.}, \bibinfo{year}{1997}.
\newblock \bibinfo{title}{{Statistics of cosmic microwave background
  polarization}}.
\newblock \bibinfo{journal}{Phys. Rev. D} \bibinfo{volume}{55},
  \bibinfo{pages}{7368--7388}.
\newblock \DOIprefix\doi{10.1103/PhysRevD.55.7368},
  \href{http://arxiv.org/abs/astro-ph/9611125}{{\tt arXiv:astro-ph/9611125}}.
\bibitem[{Kamionkowski and Kovetz(2016)}]{Kamionkowski:2015yta}
\bibinfo{author}{Kamionkowski, M.}, \bibinfo{author}{Kovetz, E.D.},
  \bibinfo{year}{2016}.
\newblock \bibinfo{title}{{The Quest for B Modes from Inflationary
  Gravitational Waves}}.
\newblock \bibinfo{journal}{Ann. Rev. Astron. Astrophys.} \bibinfo{volume}{54},
  \bibinfo{pages}{227--269}.
\newblock \DOIprefix\doi{10.1146/annurev-astro-081915-023433},
  \href{http://arxiv.org/abs/1510.06042}{{\tt arXiv:1510.06042}}.
\bibitem[{Kawamura et~al.(2021)}]{Kawamura:2020pcg}
\bibinfo{author}{Kawamura, S.}, et~al., \bibinfo{year}{2021}.
\newblock \bibinfo{title}{{Current status of space gravitational wave antenna
  DECIGO and B-DECIGO}}.
\newblock \bibinfo{journal}{PTEP} \bibinfo{volume}{2021},
  \bibinfo{pages}{05A105}.
\newblock \DOIprefix\doi{10.1093/ptep/ptab019},
  \href{http://arxiv.org/abs/2006.13545}{{\tt arXiv:2006.13545}}.
\bibitem[{Khodadi(2021)}]{Khodadi:2021owg}
\bibinfo{author}{Khodadi, M.}, \bibinfo{year}{2021}.
\newblock \bibinfo{title}{{Black Hole Superradiance in the Presence of Lorentz
  Symmetry Violation}}.
\newblock \bibinfo{journal}{Phys. Rev. D} \bibinfo{volume}{103},
  \bibinfo{pages}{064051}.
\newblock \DOIprefix\doi{10.1103/PhysRevD.103.064051},
  \href{http://arxiv.org/abs/2103.03611}{{\tt arXiv:2103.03611}}.
\bibitem[{Khodadi(2022)}]{Khodadi:2022dff}
\bibinfo{author}{Khodadi, M.}, \bibinfo{year}{2022}.
\newblock \bibinfo{title}{{Magnetic reconnection and energy extraction from a
  spinning black hole with broken Lorentz symmetry}}.
\newblock \bibinfo{journal}{Phys. Rev. D} \bibinfo{volume}{105},
  \bibinfo{pages}{023025}.
\newblock \DOIprefix\doi{10.1103/PhysRevD.105.023025},
  \href{http://arxiv.org/abs/2201.02765}{{\tt arXiv:2201.02765}}.
\bibitem[{Khodadi et~al.(2021)Khodadi, Dey and Lambiase}]{Khodadi:2021ees}
\bibinfo{author}{Khodadi, M.}, \bibinfo{author}{Dey, U.K.},
  \bibinfo{author}{Lambiase, G.}, \bibinfo{year}{2021}.
\newblock \bibinfo{title}{{Strongly magnetized hot QCD matter and stochastic
  gravitational wave background}}.
\newblock \bibinfo{journal}{Phys. Rev. D} \bibinfo{volume}{104},
  \bibinfo{pages}{063039}.
\newblock \DOIprefix\doi{10.1103/PhysRevD.104.063039},
  \href{http://arxiv.org/abs/2108.09320}{{\tt arXiv:2108.09320}}.
\bibitem[{Khodadi et~al.(2023a)Khodadi, Lambiase and
  Mastrototaro}]{Khodadi:2023yiw}
\bibinfo{author}{Khodadi, M.}, \bibinfo{author}{Lambiase, G.},
  \bibinfo{author}{Mastrototaro, L.}, \bibinfo{year}{2023}a.
\newblock \bibinfo{title}{{Spontaneous Lorentz symmetry breaking effects on
  GRBs jets arising from neutrino pair annihilation process near a black
  hole}}.
\newblock \bibinfo{journal}{Eur. Phys. J. C} \bibinfo{volume}{83},
  \bibinfo{pages}{239}.
\newblock \DOIprefix\doi{10.1140/epjc/s10052-023-11369-2},
  \href{http://arxiv.org/abs/2302.14200}{{\tt arXiv:2302.14200}}.
\bibitem[{Khodadi et~al.(2023b)Khodadi, Lambiase and Sheykhi}]{Khodadi:2022mzt}
\bibinfo{author}{Khodadi, M.}, \bibinfo{author}{Lambiase, G.},
  \bibinfo{author}{Sheykhi, A.}, \bibinfo{year}{2023}b.
\newblock \bibinfo{title}{{Constraining the Lorentz-violating bumblebee vector
  field with big bang nucleosynthesis and gravitational baryogenesis}}.
\newblock \bibinfo{journal}{Eur. Phys. J. C} \bibinfo{volume}{83},
  \bibinfo{pages}{386}.
\newblock \DOIprefix\doi{10.1140/epjc/s10052-023-11546-3},
  \href{http://arxiv.org/abs/2211.07934}{{\tt arXiv:2211.07934}}.
\bibitem[{Khodadi et~al.(2018)Khodadi, Nozari, Abedi and
  Capozziello}]{Khodadi:2018scn}
\bibinfo{author}{Khodadi, M.}, \bibinfo{author}{Nozari, K.},
  \bibinfo{author}{Abedi, H.}, \bibinfo{author}{Capozziello, S.},
  \bibinfo{year}{2018}.
\newblock \bibinfo{title}{{Planck scale effects on the stochastic gravitational
  wave background generated from cosmological hadronization transition: A
  qualitative study}}.
\newblock \bibinfo{journal}{Phys. Lett. B} \bibinfo{volume}{783},
  \bibinfo{pages}{326--333}.
\newblock \DOIprefix\doi{10.1016/j.physletb.2018.07.010},
  \href{http://arxiv.org/abs/1805.11310}{{\tt arXiv:1805.11310}}.
\bibitem[{Kostelecky and Tasson(2011)}]{Kostelecky:2010ze}
\bibinfo{author}{Kostelecky, A.V.}, \bibinfo{author}{Tasson, J.D.},
  \bibinfo{year}{2011}.
\newblock \bibinfo{title}{{Matter-gravity couplings and Lorentz violation}}.
\newblock \bibinfo{journal}{Phys. Rev. D} \bibinfo{volume}{83},
  \bibinfo{pages}{016013}.
\newblock \DOIprefix\doi{10.1103/PhysRevD.83.016013},
  \href{http://arxiv.org/abs/1006.4106}{{\tt arXiv:1006.4106}}.
\bibitem[{Kostelecky(2004)}]{Kostelecky:2003fs}
\bibinfo{author}{Kostelecky, V.A.}, \bibinfo{year}{2004}.
\newblock \bibinfo{title}{{Gravity, Lorentz violation, and the standard
  model}}.
\newblock \bibinfo{journal}{Phys. Rev. D} \bibinfo{volume}{69},
  \bibinfo{pages}{105009}.
\newblock \DOIprefix\doi{10.1103/PhysRevD.69.105009},
  \href{http://arxiv.org/abs/hep-th/0312310}{{\tt arXiv:hep-th/0312310}}.
\bibitem[{Kostelecky and Samuel(1989a)}]{Kostelecky:1989jw}
\bibinfo{author}{Kostelecky, V.A.}, \bibinfo{author}{Samuel, S.},
  \bibinfo{year}{1989}a.
\newblock \bibinfo{title}{{Gravitational Phenomenology in Higher Dimensional
  Theories and Strings}}.
\newblock \bibinfo{journal}{Phys. Rev. D} \bibinfo{volume}{40},
  \bibinfo{pages}{1886--1903}.
\newblock \DOIprefix\doi{10.1103/PhysRevD.40.1886}.
\bibitem[{Kostelecky and Samuel(1989b)}]{Kostelecky:1988zi}
\bibinfo{author}{Kostelecky, V.A.}, \bibinfo{author}{Samuel, S.},
  \bibinfo{year}{1989}b.
\newblock \bibinfo{title}{{Spontaneous Breaking of Lorentz Symmetry in String
  Theory}}.
\newblock \bibinfo{journal}{Phys. Rev. D} \bibinfo{volume}{39},
  \bibinfo{pages}{683}.
\newblock \DOIprefix\doi{10.1103/PhysRevD.39.683}.
\bibitem[{Krishnan et~al.(2021)Krishnan, Mohayaee, Colg\'ain, Sheikh-Jabbari
  and Yin}]{Krishnan:2021dyb}
\bibinfo{author}{Krishnan, C.}, \bibinfo{author}{Mohayaee, R.},
  \bibinfo{author}{Colg\'ain, E.O.}, \bibinfo{author}{Sheikh-Jabbari, M.M.},
  \bibinfo{author}{Yin, L.}, \bibinfo{year}{2021}.
\newblock \bibinfo{title}{{Does Hubble tension signal a breakdown in FLRW
  cosmology?}}
\newblock \bibinfo{journal}{Class. Quant. Grav.} \bibinfo{volume}{38},
  \bibinfo{pages}{184001}.
\newblock \DOIprefix\doi{10.1088/1361-6382/ac1a81},
  \href{http://arxiv.org/abs/2105.09790}{{\tt arXiv:2105.09790}}.
\bibitem[{Krishnan et~al.(2022)Krishnan, Mohayaee, Colg\'ain, Sheikh-Jabbari
  and Yin}]{Krishnan:2021jmh}
\bibinfo{author}{Krishnan, C.}, \bibinfo{author}{Mohayaee, R.},
  \bibinfo{author}{Colg\'ain, E.O.}, \bibinfo{author}{Sheikh-Jabbari, M.M.},
  \bibinfo{author}{Yin, L.}, \bibinfo{year}{2022}.
\newblock \bibinfo{title}{{Hints of FLRW breakdown from supernovae}}.
\newblock \bibinfo{journal}{Phys. Rev. D} \bibinfo{volume}{105},
  \bibinfo{pages}{063514}.
\newblock \DOIprefix\doi{10.1103/PhysRevD.105.063514},
  \href{http://arxiv.org/abs/2106.02532}{{\tt arXiv:2106.02532}}.
\bibitem[{Leitao and Megevand(2016)}]{Leitao:2015fmj}
\bibinfo{author}{Leitao, L.}, \bibinfo{author}{Megevand, A.},
  \bibinfo{year}{2016}.
\newblock \bibinfo{title}{{Gravitational waves from a very strong electroweak
  phase transition}}.
\newblock \bibinfo{journal}{JCAP} \bibinfo{volume}{05}, \bibinfo{pages}{037}.
\newblock \DOIprefix\doi{10.1088/1475-7516/2016/05/037},
  \href{http://arxiv.org/abs/1512.08962}{{\tt arXiv:1512.08962}}.
\bibitem[{Li et~al.(2018)Li, Yang and Yuan}]{Li:2018oqf}
\bibinfo{author}{Li, M.W.}, \bibinfo{author}{Yang, Y.}, \bibinfo{author}{Yuan,
  P.H.}, \bibinfo{year}{2018}.
\newblock \bibinfo{title}{{Imprints of Early Universe on Gravitational Waves
  from First-Order Phase Transition in QCD}}
  \href{http://arxiv.org/abs/1812.09676}{{\tt arXiv:1812.09676}}.
\bibitem[{Liu et~al.(2024)Liu, Guo, Wei and Liu}]{Liu:2024axg}
\bibinfo{author}{Liu, J.Z.}, \bibinfo{author}{Guo, W.D.}, \bibinfo{author}{Wei,
  S.W.}, \bibinfo{author}{Liu, Y.X.}, \bibinfo{year}{2024}.
\newblock \bibinfo{title}{{Charged spherically symmetric and slowly rotating
  charged black hole solutions in bumblebee gravity}}
  \href{http://arxiv.org/abs/2407.08396}{{\tt arXiv:2407.08396}}.
\bibitem[{Maggiore(2000)}]{Maggiore:1999vm}
\bibinfo{author}{Maggiore, M.}, \bibinfo{year}{2000}.
\newblock \bibinfo{title}{{Gravitational wave experiments and early universe
  cosmology}}.
\newblock \bibinfo{journal}{Phys. Rept.} \bibinfo{volume}{331},
  \bibinfo{pages}{283--367}.
\newblock \DOIprefix\doi{10.1016/S0370-1573(99)00102-7},
  \href{http://arxiv.org/abs/gr-qc/9909001}{{\tt arXiv:gr-qc/9909001}}.
\bibitem[{Mai et~al.(2024)Mai, Xu, Liang and Shao}]{Mai:2024lgk}
\bibinfo{author}{Mai, Z.F.}, \bibinfo{author}{Xu, R.}, \bibinfo{author}{Liang,
  D.}, \bibinfo{author}{Shao, L.}, \bibinfo{year}{2024}.
\newblock \bibinfo{title}{{Dynamic instability analysis for bumblebee black
  holes: The odd parity}}.
\newblock \bibinfo{journal}{Phys. Rev. D} \bibinfo{volume}{109},
  \bibinfo{pages}{084076}.
\newblock \DOIprefix\doi{10.1103/PhysRevD.109.084076},
  \href{http://arxiv.org/abs/2401.07757}{{\tt arXiv:2401.07757}}.
\bibitem[{Maleknejad and Sheikh-Jabbari(2012)}]{Maleknejad:2012as}
\bibinfo{author}{Maleknejad, A.}, \bibinfo{author}{Sheikh-Jabbari, M.M.},
  \bibinfo{year}{2012}.
\newblock \bibinfo{title}{{Revisiting Cosmic No-Hair Theorem for Inflationary
  Settings}}.
\newblock \bibinfo{journal}{Phys. Rev. D} \bibinfo{volume}{85},
  \bibinfo{pages}{123508}.
\newblock \DOIprefix\doi{10.1103/PhysRevD.85.123508},
  \href{http://arxiv.org/abs/1203.0219}{{\tt arXiv:1203.0219}}.
\bibitem[{Maluf and Neves(2021a)}]{Maluf:2020kgf}
\bibinfo{author}{Maluf, R.V.}, \bibinfo{author}{Neves, J.C.S.},
  \bibinfo{year}{2021}a.
\newblock \bibinfo{title}{{Black holes with a cosmological constant in
  bumblebee gravity}}.
\newblock \bibinfo{journal}{Phys. Rev. D} \bibinfo{volume}{103},
  \bibinfo{pages}{044002}.
\newblock \DOIprefix\doi{10.1103/PhysRevD.103.044002},
  \href{http://arxiv.org/abs/2011.12841}{{\tt arXiv:2011.12841}}.
\bibitem[{Maluf and Neves(2021b)}]{Maluf:2021lwh}
\bibinfo{author}{Maluf, R.V.}, \bibinfo{author}{Neves, J.C.S.},
  \bibinfo{year}{2021}b.
\newblock \bibinfo{title}{{Bumblebee field as a source of cosmological
  anisotropies}}.
\newblock \bibinfo{journal}{JCAP} \bibinfo{volume}{10}, \bibinfo{pages}{038}.
\newblock \DOIprefix\doi{10.1088/1475-7516/2021/10/038},
  \href{http://arxiv.org/abs/2105.08659}{{\tt arXiv:2105.08659}}.
\bibitem[{Mariz et~al.(2023)Mariz, Nascimento and Petrov}]{Mariz:2022oib}
\bibinfo{author}{Mariz, T.}, \bibinfo{author}{Nascimento, J.R.},
  \bibinfo{author}{Petrov, A.}, \bibinfo{year}{2023}.
\newblock \bibinfo{title}{{Lorentz Symmetry Breaking\textemdash{}Classical and
  Quantum Aspects}}.
\newblock SpringerBriefs in Physics, \bibinfo{publisher}{Springer}.
\newblock \DOIprefix\doi{10.1007/978-3-031-20120-2},
  \href{http://arxiv.org/abs/2205.02594}{{\tt arXiv:2205.02594}}.
\bibitem[{Mavromatos et~al.(2022)Mavromatos, Spanos and
  Stamou}]{Mavromatos:2022yql}
\bibinfo{author}{Mavromatos, N.E.}, \bibinfo{author}{Spanos, V.C.},
  \bibinfo{author}{Stamou, I.D.}, \bibinfo{year}{2022}.
\newblock \bibinfo{title}{{Primordial black holes and gravitational waves in
  multiaxion-Chern-Simons inflation}}.
\newblock \bibinfo{journal}{Phys. Rev. D} \bibinfo{volume}{106},
  \bibinfo{pages}{063532}.
\newblock \DOIprefix\doi{10.1103/PhysRevD.106.063532},
  \href{http://arxiv.org/abs/2206.07963}{{\tt arXiv:2206.07963}}.
\bibitem[{Meijer et~al.(2024)Meijer, Lopez, Tsuna and Caudill}]{Meijer:2023yhn}
\bibinfo{author}{Meijer, Q.}, \bibinfo{author}{Lopez, M.},
  \bibinfo{author}{Tsuna, D.}, \bibinfo{author}{Caudill, S.},
  \bibinfo{year}{2024}.
\newblock \bibinfo{title}{{Gravitational-wave searches for cosmic string cusps
  in Einstein Telescope data using deep learning}}.
\newblock \bibinfo{journal}{Phys. Rev. D} \bibinfo{volume}{109},
  \bibinfo{pages}{022006}.
\newblock \DOIprefix\doi{10.1103/PhysRevD.109.022006},
  \href{http://arxiv.org/abs/2308.12323}{{\tt arXiv:2308.12323}}.
\bibitem[{Mohamadnejad(2022)}]{Mohamadnejad:2021tke}
\bibinfo{author}{Mohamadnejad, A.}, \bibinfo{year}{2022}.
\newblock \bibinfo{title}{{Electroweak phase transition and gravitational waves
  in a two-component dark matter model}}.
\newblock \bibinfo{journal}{JHEP} \bibinfo{volume}{03}, \bibinfo{pages}{188}.
\newblock \DOIprefix\doi{10.1007/JHEP03(2022)188},
  \href{http://arxiv.org/abs/2111.04342}{{\tt arXiv:2111.04342}}.
\bibitem[{Mukhanov et~al.(1992)Mukhanov, Feldman and
  Brandenberger}]{Mukhanov:1990me}
\bibinfo{author}{Mukhanov, V.F.}, \bibinfo{author}{Feldman, H.A.},
  \bibinfo{author}{Brandenberger, R.H.}, \bibinfo{year}{1992}.
\newblock \bibinfo{title}{{Theory of cosmological perturbations. Part 1.
  Classical perturbations. Part 2. Quantum theory of perturbations. Part 3.
  Extensions}}.
\newblock \bibinfo{journal}{Phys. Rept.} \bibinfo{volume}{215},
  \bibinfo{pages}{203--333}.
\newblock \DOIprefix\doi{10.1016/0370-1573(92)90044-Z}.
\bibitem[{Neves(2023)}]{Neves:2022qyb}
\bibinfo{author}{Neves, J.C.S.}, \bibinfo{year}{2023}.
\newblock \bibinfo{title}{{Kasner cosmology in bumblebee gravity}}.
\newblock \bibinfo{journal}{Annals Phys.} \bibinfo{volume}{454},
  \bibinfo{pages}{169338}.
\newblock \DOIprefix\doi{10.1016/j.aop.2023.169338},
  \href{http://arxiv.org/abs/2209.00589}{{\tt arXiv:2209.00589}}.
\bibitem[{Neves and Gardim(2024)}]{Neves:2024ggn}
\bibinfo{author}{Neves, J.C.S.}, \bibinfo{author}{Gardim, F.G.},
  \bibinfo{year}{2024}.
\newblock \bibinfo{title}{{Stars and quark stars in bumblebee gravity}}
  \href{http://arxiv.org/abs/2409.20360}{{\tt arXiv:2409.20360}}.
\bibitem[{Ni(2013)}]{Ni:2012eh}
\bibinfo{author}{Ni, W.T.}, \bibinfo{year}{2013}.
\newblock \bibinfo{title}{{ASTROD-GW: Overview and Progress}}.
\newblock \bibinfo{journal}{Int. J. Mod. Phys. D} \bibinfo{volume}{22},
  \bibinfo{pages}{1341004}.
\newblock \DOIprefix\doi{10.1142/S0218271813410046},
  \href{http://arxiv.org/abs/1212.2816}{{\tt arXiv:1212.2816}}.
\bibitem[{Nilsson(2022)}]{Nilsson:2022mzq}
\bibinfo{author}{Nilsson, N.A.}, \bibinfo{year}{2022}.
\newblock \bibinfo{title}{{Explicit spacetime-symmetry breaking and the
  dynamics of primordial fields}}.
\newblock \bibinfo{journal}{Phys. Rev. D} \bibinfo{volume}{106},
  \bibinfo{pages}{104036}.
\newblock \DOIprefix\doi{10.1103/PhysRevD.106.104036},
  \href{http://arxiv.org/abs/2205.00496}{{\tt arXiv:2205.00496}}.
\bibitem[{Odintsov et~al.(2024)Odintsov, D'Onofrio and Paul}]{Odintsov:2024sbo}
\bibinfo{author}{Odintsov, S.D.}, \bibinfo{author}{D'Onofrio, S.},
  \bibinfo{author}{Paul, T.}, \bibinfo{year}{2024}.
\newblock \bibinfo{title}{{Primordial gravitational waves in horizon cosmology
  and constraints on entropic parameters}}.
\newblock \bibinfo{journal}{Phys. Rev. D} \bibinfo{volume}{110},
  \bibinfo{pages}{043539}.
\newblock \DOIprefix\doi{10.1103/PhysRevD.110.043539},
  \href{http://arxiv.org/abs/2407.05855}{{\tt arXiv:2407.05855}}.
\bibitem[{Odintsov et~al.(2022)Odintsov, Oikonomou and
  Myrzakulov}]{Odintsov:2022cbm}
\bibinfo{author}{Odintsov, S.D.}, \bibinfo{author}{Oikonomou, V.K.},
  \bibinfo{author}{Myrzakulov, R.}, \bibinfo{year}{2022}.
\newblock \bibinfo{title}{{Spectrum of Primordial Gravitational Waves in
  Modified Gravities: A Short Overview}}.
\newblock \bibinfo{journal}{Symmetry} \bibinfo{volume}{14},
  \bibinfo{pages}{729}.
\newblock \DOIprefix\doi{10.3390/sym14040729},
  \href{http://arxiv.org/abs/2204.00876}{{\tt arXiv:2204.00876}}.
\bibitem[{Oikonomou(2023)}]{Oikonomou:2022ijs}
\bibinfo{author}{Oikonomou, V.K.}, \bibinfo{year}{2023}.
\newblock \bibinfo{title}{{Amplification of the Primordial Gravitational Waves
  Energy Spectrum by a Kinetic Scalar in $F(R)$ Gravity}}.
\newblock \bibinfo{journal}{Astropart. Phys.} \bibinfo{volume}{144},
  \bibinfo{pages}{102777}.
\newblock \DOIprefix\doi{10.1016/j.astropartphys.2022.102777},
  \href{http://arxiv.org/abs/2209.09781}{{\tt arXiv:2209.09781}}.
\bibitem[{\"Ovg\"un et~al.(2019)\"Ovg\"un, Jusufi and
  Sakall\i{}}]{Ovgun:2018xys}
\bibinfo{author}{\"Ovg\"un, A.}, \bibinfo{author}{Jusufi, K.},
  \bibinfo{author}{Sakall\i{}, I.}, \bibinfo{year}{2019}.
\newblock \bibinfo{title}{{Exact traversable wormhole solution in bumblebee
  gravity}}.
\newblock \bibinfo{journal}{Phys. Rev. D} \bibinfo{volume}{99},
  \bibinfo{pages}{024042}.
\newblock \DOIprefix\doi{10.1103/PhysRevD.99.024042},
  \href{http://arxiv.org/abs/1804.09911}{{\tt arXiv:1804.09911}}.
\bibitem[{P\'aramos and Guiomar(2014)}]{Paramos:2014mda}
\bibinfo{author}{P\'aramos, J.}, \bibinfo{author}{Guiomar, G.},
  \bibinfo{year}{2014}.
\newblock \bibinfo{title}{{Astrophysical Constraints on the Bumblebee Model}}.
\newblock \bibinfo{journal}{Phys. Rev. D} \bibinfo{volume}{90},
  \bibinfo{pages}{082002}.
\newblock \DOIprefix\doi{10.1103/PhysRevD.90.082002},
  \href{http://arxiv.org/abs/1409.2022}{{\tt arXiv:1409.2022}}.
\bibitem[{Pavlopoulos(1967)}]{Pavlopoulos:1967dm}
\bibinfo{author}{Pavlopoulos, T.G.}, \bibinfo{year}{1967}.
\newblock \bibinfo{title}{{Breakdown of Lorentz invariance}}.
\newblock \bibinfo{journal}{Phys. Rev.} \bibinfo{volume}{159},
  \bibinfo{pages}{1106--1110}.
\newblock \DOIprefix\doi{10.1103/PhysRev.159.1106}.
\bibitem[{Rezapour et~al.(2022)Rezapour, Bitaghsir~Fadafan and
  Ahmadvand}]{Rezapour:2020mvi}
\bibinfo{author}{Rezapour, S.}, \bibinfo{author}{Bitaghsir~Fadafan, K.},
  \bibinfo{author}{Ahmadvand, M.}, \bibinfo{year}{2022}.
\newblock \bibinfo{title}{{Gravitational waves of a first-order QCD phase
  transition at finite coupling from holography}}.
\newblock \bibinfo{journal}{Annals Phys.} \bibinfo{volume}{437},
  \bibinfo{pages}{168731}.
\newblock \DOIprefix\doi{10.1016/j.aop.2021.168731},
  \href{http://arxiv.org/abs/2006.04265}{{\tt arXiv:2006.04265}}.
\bibitem[{Santos et~al.(2015)Santos, Petrov, Jesus and
  Nascimento}]{Santos:2014nxm}
\bibinfo{author}{Santos, A.F.}, \bibinfo{author}{Petrov, A.Y.},
  \bibinfo{author}{Jesus, W.D.R.}, \bibinfo{author}{Nascimento, J.R.},
  \bibinfo{year}{2015}.
\newblock \bibinfo{title}{{G\"odel solution in the bumblebee gravity}}.
\newblock \bibinfo{journal}{Mod. Phys. Lett. A} \bibinfo{volume}{30},
  \bibinfo{pages}{1550011}.
\newblock \DOIprefix\doi{10.1142/S021773231550011X},
  \href{http://arxiv.org/abs/1407.5985}{{\tt arXiv:1407.5985}}.
\bibitem[{Sarmah and Goswami(2024)}]{Sarmah:2024xwx}
\bibinfo{author}{Sarmah, P.}, \bibinfo{author}{Goswami, U.D.},
  \bibinfo{year}{2024}.
\newblock \bibinfo{title}{{Anisotropic cosmology in Bumblebee gravity theory}}
  \href{http://arxiv.org/abs/2407.13487}{{\tt arXiv:2407.13487}}.
\bibitem[{Sasaki et~al.(2018)Sasaki, Suyama, Tanaka and
  Yokoyama}]{Sasaki:2018dmp}
\bibinfo{author}{Sasaki, M.}, \bibinfo{author}{Suyama, T.},
  \bibinfo{author}{Tanaka, T.}, \bibinfo{author}{Yokoyama, S.},
  \bibinfo{year}{2018}.
\newblock \bibinfo{title}{{Primordial black holes\textemdash{}perspectives in
  gravitational wave astronomy}}.
\newblock \bibinfo{journal}{Class. Quant. Grav.} \bibinfo{volume}{35},
  \bibinfo{pages}{063001}.
\newblock \DOIprefix\doi{10.1088/1361-6382/aaa7b4},
  \href{http://arxiv.org/abs/1801.05235}{{\tt arXiv:1801.05235}}.
\bibitem[{Sathyaprakash et~al.(2012)}]{Sathyaprakash:2012jk}
\bibinfo{author}{Sathyaprakash, B.}, et~al., \bibinfo{year}{2012}.
\newblock \bibinfo{title}{{Scientific Objectives of Einstein Telescope}}.
\newblock \bibinfo{journal}{Class. Quant. Grav.} \bibinfo{volume}{29},
  \bibinfo{pages}{124013}.
\newblock \DOIprefix\doi{10.1088/0264-9381/29/12/124013},
  \href{http://arxiv.org/abs/1206.0331}{{\tt arXiv:1206.0331}}.
  \bibinfo{note}{[Erratum: Class.Quant.Grav. 30, 079501 (2013)]}.
\bibitem[{Secrest et~al.(2022)Secrest, von Hausegger, Rameez, Mohayaee and
  Sarkar}]{Secrest:2022uvx}
\bibinfo{author}{Secrest, N.J.}, \bibinfo{author}{von Hausegger, S.},
  \bibinfo{author}{Rameez, M.}, \bibinfo{author}{Mohayaee, R.},
  \bibinfo{author}{Sarkar, S.}, \bibinfo{year}{2022}.
\newblock \bibinfo{title}{{A Challenge to the Standard Cosmological Model}}.
\newblock \bibinfo{journal}{Astrophys. J. Lett.} \bibinfo{volume}{937},
  \bibinfo{pages}{L31}.
\newblock \DOIprefix\doi{10.3847/2041-8213/ac88c0},
  \href{http://arxiv.org/abs/2206.05624}{{\tt arXiv:2206.05624}}.
\bibitem[{Sesana et~al.(2021)}]{Sesana:2019vho}
\bibinfo{author}{Sesana, A.}, et~al., \bibinfo{year}{2021}.
\newblock \bibinfo{title}{{Unveiling the gravitational universe at $\mu$-Hz
  frequencies}}.
\newblock \bibinfo{journal}{Exper. Astron.} \bibinfo{volume}{51},
  \bibinfo{pages}{1333--1383}.
\newblock \DOIprefix\doi{10.1007/s10686-021-09709-9},
  \href{http://arxiv.org/abs/1908.11391}{{\tt arXiv:1908.11391}}.
\bibitem[{Shajiee and Tofighi(2019)}]{Shajiee:2018jdq}
\bibinfo{author}{Shajiee, V.R.}, \bibinfo{author}{Tofighi, A.},
  \bibinfo{year}{2019}.
\newblock \bibinfo{title}{{Electroweak Phase Transition, Gravitational Waves
  and Dark Matter in Two Scalar Singlet Extension of The Standard Model}}.
\newblock \bibinfo{journal}{Eur. Phys. J. C} \bibinfo{volume}{79},
  \bibinfo{pages}{360}.
\newblock \DOIprefix\doi{10.1140/epjc/s10052-019-6881-6},
  \href{http://arxiv.org/abs/1811.09807}{{\tt arXiv:1811.09807}}.
\bibitem[{Shakeri and Allahyari(2018)}]{Shakeri:2018qal}
\bibinfo{author}{Shakeri, S.}, \bibinfo{author}{Allahyari, A.},
  \bibinfo{year}{2018}.
\newblock \bibinfo{title}{{Circularly Polarized EM Radiation from GW Binary
  Sources}}.
\newblock \bibinfo{journal}{JCAP} \bibinfo{volume}{11}, \bibinfo{pages}{042}.
\newblock \DOIprefix\doi{10.1088/1475-7516/2018/11/042},
  \href{http://arxiv.org/abs/1808.05210}{{\tt arXiv:1808.05210}}.
\bibitem[{Sousa(2024)}]{Sousa:2024ytl}
\bibinfo{author}{Sousa, L.}, \bibinfo{year}{2024}.
\newblock \bibinfo{title}{{Cosmic strings and gravitational waves}}.
\newblock \bibinfo{journal}{Gen. Rel. Grav.} \bibinfo{volume}{56},
  \bibinfo{pages}{105}.
\newblock \DOIprefix\doi{10.1007/s10714-024-03293-x}.
\bibitem[{Tristram et~al.(2022)}]{Tristram:2021tvh}
\bibinfo{author}{Tristram, M.}, et~al., \bibinfo{year}{2022}.
\newblock \bibinfo{title}{{Improved limits on the tensor-to-scalar ratio using
  BICEP and Planck data}}.
\newblock \bibinfo{journal}{Phys. Rev. D} \bibinfo{volume}{105},
  \bibinfo{pages}{083524}.
\newblock \DOIprefix\doi{10.1103/PhysRevD.105.083524},
  \href{http://arxiv.org/abs/2112.07961}{{\tt arXiv:2112.07961}}.
\bibitem[{Turner(1997)}]{Turner:1996ck}
\bibinfo{author}{Turner, M.S.}, \bibinfo{year}{1997}.
\newblock \bibinfo{title}{{Detectability of inflation produced gravitational
  waves}}.
\newblock \bibinfo{journal}{Phys. Rev. D} \bibinfo{volume}{55},
  \bibinfo{pages}{R435--R439}.
\newblock \DOIprefix\doi{10.1103/PhysRevD.55.R435},
  \href{http://arxiv.org/abs/astro-ph/9607066}{{\tt arXiv:astro-ph/9607066}}.
\bibitem[{Vagnozzi(2023)}]{Vagnozzi:2023lwo}
\bibinfo{author}{Vagnozzi, S.}, \bibinfo{year}{2023}.
\newblock \bibinfo{title}{{Inflationary interpretation of the stochastic
  gravitational wave background signal detected by pulsar timing array
  experiments}}.
\newblock \bibinfo{journal}{JHEAp} \bibinfo{volume}{39},
  \bibinfo{pages}{81--98}.
\newblock \DOIprefix\doi{10.1016/j.jheap.2023.07.001},
  \href{http://arxiv.org/abs/2306.16912}{{\tt arXiv:2306.16912}}.
\bibitem[{Vagnozzi et~al.(2023)}]{Vagnozzi:2022moj}
\bibinfo{author}{Vagnozzi, S.}, et~al., \bibinfo{year}{2023}.
\newblock \bibinfo{title}{{Horizon-scale tests of gravity theories and
  fundamental physics from the Event Horizon Telescope image of Sagittarius
  A}}.
\newblock \bibinfo{journal}{Class. Quant. Grav.} \bibinfo{volume}{40},
  \bibinfo{pages}{165007}.
\newblock \DOIprefix\doi{10.1088/1361-6382/acd97b},
  \href{http://arxiv.org/abs/2205.07787}{{\tt arXiv:2205.07787}}.
\bibitem[{Valtancoli(2023)}]{Valtancoli:2023kdy}
\bibinfo{author}{Valtancoli, P.}, \bibinfo{year}{2023}.
\newblock \bibinfo{title}{{Bumblebee gravity with cosmological constant}}
  \href{http://arxiv.org/abs/2308.05328}{{\tt arXiv:2308.05328}}.
\bibitem[{Wald(1983)}]{Wald:1983ky}
\bibinfo{author}{Wald, R.M.}, \bibinfo{year}{1983}.
\newblock \bibinfo{title}{{Asymptotic behavior of homogeneous cosmological
  models in the presence of a positive cosmological constant}}.
\newblock \bibinfo{journal}{Phys. Rev. D} \bibinfo{volume}{28},
  \bibinfo{pages}{2118--2120}.
\newblock \DOIprefix\doi{10.1103/PhysRevD.28.2118}.
\bibitem[{Wang(2024)}]{Wang:2024gko}
\bibinfo{author}{Wang, D.}, \bibinfo{year}{2024}.
\newblock \bibinfo{title}{{Primordial Gravitational Waves 2024}}
  \href{http://arxiv.org/abs/2407.02714}{{\tt arXiv:2407.02714}}.
\bibitem[{Wang et~al.(2022)Wang, Chen and Jing}]{Wang:2021gtd}
\bibinfo{author}{Wang, Z.}, \bibinfo{author}{Chen, S.}, \bibinfo{author}{Jing,
  J.}, \bibinfo{year}{2022}.
\newblock \bibinfo{title}{{Constraint on parameters of a rotating black hole in
  Einstein-bumblebee theory by quasi-periodic oscillations}}.
\newblock \bibinfo{journal}{Eur. Phys. J. C} \bibinfo{volume}{82},
  \bibinfo{pages}{528}.
\newblock \DOIprefix\doi{10.1140/epjc/s10052-022-10475-x},
  \href{http://arxiv.org/abs/2112.02895}{{\tt arXiv:2112.02895}}.
\bibitem[{Watanabe and Komatsu(2006)}]{Watanabe:2006qe}
\bibinfo{author}{Watanabe, Y.}, \bibinfo{author}{Komatsu, E.},
  \bibinfo{year}{2006}.
\newblock \bibinfo{title}{{Improved Calculation of the Primordial Gravitational
  Wave Spectrum in the Standard Model}}.
\newblock \bibinfo{journal}{Phys. Rev. D} \bibinfo{volume}{73},
  \bibinfo{pages}{123515}.
\newblock \DOIprefix\doi{10.1103/PhysRevD.73.123515},
  \href{http://arxiv.org/abs/astro-ph/0604176}{{\tt arXiv:astro-ph/0604176}}.
\bibitem[{Weinberg(2004)}]{Weinberg:2003ur}
\bibinfo{author}{Weinberg, S.}, \bibinfo{year}{2004}.
\newblock \bibinfo{title}{{Damping of tensor modes in cosmology}}.
\newblock \bibinfo{journal}{Phys. Rev. D} \bibinfo{volume}{69},
  \bibinfo{pages}{023503}.
\newblock \DOIprefix\doi{10.1103/PhysRevD.69.023503},
  \href{http://arxiv.org/abs/astro-ph/0306304}{{\tt arXiv:astro-ph/0306304}}.
\bibitem[{Weir(2018)}]{Weir:2017wfa}
\bibinfo{author}{Weir, D.J.}, \bibinfo{year}{2018}.
\newblock \bibinfo{title}{{Gravitational waves from a first order electroweak
  phase transition: a brief review}}.
\newblock \bibinfo{journal}{Phil. Trans. Roy. Soc. Lond. A}
  \bibinfo{volume}{376}, \bibinfo{pages}{20170126}.
\newblock \DOIprefix\doi{10.1098/rsta.2017.0126},
  \href{http://arxiv.org/abs/1705.01783}{{\tt arXiv:1705.01783}}.
  \bibinfo{note}{[Erratum: Phil.Trans.Roy.Soc.Lond.A 381, 20230212 (2023)]}.
\bibitem[{Xu et~al.(2023)Xu, Liang and Shao}]{Xu:2022frb}
\bibinfo{author}{Xu, R.}, \bibinfo{author}{Liang, D.}, \bibinfo{author}{Shao,
  L.}, \bibinfo{year}{2023}.
\newblock \bibinfo{title}{{Static spherical vacuum solutions in the bumblebee
  gravity model}}.
\newblock \bibinfo{journal}{Phys. Rev. D} \bibinfo{volume}{107},
  \bibinfo{pages}{024011}.
\newblock \DOIprefix\doi{10.1103/PhysRevD.107.024011},
  \href{http://arxiv.org/abs/2209.02209}{{\tt arXiv:2209.02209}}.
\bibitem[{Yonemaru et~al.(2021)}]{Yonemaru:2020bmr}
\bibinfo{author}{Yonemaru, N.}, et~al., \bibinfo{year}{2021}.
\newblock \bibinfo{title}{{Searching for gravitational wave bursts from cosmic
  string cusps with the Parkes Pulsar Timing Array}}.
\newblock \bibinfo{journal}{Mon. Not. Roy. Astron. Soc.} \bibinfo{volume}{501},
  \bibinfo{pages}{701--712}.
\newblock \DOIprefix\doi{10.1093/mnras/staa3721},
  \href{http://arxiv.org/abs/2011.13490}{{\tt arXiv:2011.13490}}.
\bibitem[{Zaldarriaga and Seljak(1997)}]{Zaldarriaga:1996xe}
\bibinfo{author}{Zaldarriaga, M.}, \bibinfo{author}{Seljak, U.},
  \bibinfo{year}{1997}.
\newblock \bibinfo{title}{{An all sky analysis of polarization in the microwave
  background}}.
\newblock \bibinfo{journal}{Phys. Rev. D} \bibinfo{volume}{55},
  \bibinfo{pages}{1830--1840}.
\newblock \DOIprefix\doi{10.1103/PhysRevD.55.1830},
  \href{http://arxiv.org/abs/astro-ph/9609170}{{\tt arXiv:astro-ph/9609170}}.

\end{thebibliography}






\end{document}